\documentstyle[pre,aps,epsfig,epsf]{revtex}
\newcommand{\institute}[1]{\parbox{16cm}{%
\centering\normalsize \sl #1}}
\newcommand{\be}{\begin{equation}}
\newcommand{\ee}{\end{equation}}
\newcommand{\bea}{\begin{eqnarray}}
\newcommand{\eea}{\end{eqnarray}}
\newcommand{\bean}{\begin{eqnarray*}}
\newcommand{\eean}{\end{eqnarray*}}
\newcommand{\uts}[2]{#1^{{\scriptsize \mbox{#2}}}}

\newcommand{\Tr}{\mbox{Tr\,}}
\newcommand{\tr}[1]{\mbox{Tr}\left(#1\right)}

\newcommand{\dtt}{{\scriptsize \mbox{d}}}
\newcommand{\Exp}{\mbox{e}}
\newcommand{\hf}[1]{#1_{x}}
\newcommand{\hfup}[2]{#1_{x+\hat{#2}}}

\newcommand{\gf}[2]{#1_{x,#2}}
\newcommand{\gfup}[3]{#1_{x+\hat{#2},#3}}

\newcommand{\lkop}[3]{#1_{#2;#3}}
\newcommand{\lphi}{L_{\varphi}}
\newcommand{\lphix}{L_{\varphi,x}}

\newcommand{\lphisx}{L_{\varphi,s,x}}
\newcommand{\lphitx}{L_{\varphi,t,x}}

\newcommand{\plaqx}{P_{{\scriptsize \mbox{p}},x}}

\newcommand{\plaqsx}{P_{{\scriptsize \mbox{p}},s,x}}
\newcommand{\plaqtx}{P_{{\scriptsize \mbox{p}},t,x}}
\newcommand{\mh}{m_H}
\newcommand{\mw}{m_W}
\newcommand{\mhtr}{m_{H,0}}
\newcommand{\mwtr}{m_{W,0}}
\newcommand{\mhs}{m_{H,s}}
\newcommand{\mws}{m_{W,s}}
\newcommand{\mht}{m_{H,t}}
\newcommand{\mwt}{m_{W,t}}

\newcommand{\msst}{m_{V,st}}
\newcommand{\msts}{m_{V,ts}}
\newcommand{\msss}{m_{V,ss}}
\newcommand{\hw}{R_{HW}}
\newcommand{\hwtr}{R_{HW,0}}
\newcommand{\hws}{R_{HW,s}}
\newcommand{\hwt}{R_{HW,t}}

\newcommand{\gb}{\gamma_{\beta}}
\newcommand{\gk}{\gamma_{\kappa}}

\newcommand{\xh}{\xi_H}
\newcommand{\xw}{\xi_W}
\newcommand{\xp}{\xi_V}
\newcommand{\Wij}{W_{ij}}
\newcommand{\Wst}{W_{st}}
\newcommand{\Wts}{W_{ts}}
\newcommand{\Wss}{W_{ss}}
\newcommand{\Vij}{V_{ij}}
\newcommand{\Vcij}{V_{{\scriptsize \mbox{cont}},ij}}
\newcommand{\Vst}{V_{st}}
\newcommand{\Vts}{V_{ts}}

\newcommand{\Vcst}{V_{{\scriptsize \mbox{cont}},st}}
\newcommand{\Vcts}{V_{{\scriptsize \mbox{cont}},ts}}

\newcommand{\aij}{A_{ij}}
\newcommand{\cij}{C_{ij}}
\newcommand{\dij}{D_{ij}}
\newcommand{\gij}{G_{ij}}
\newcommand{\iij}{I_{ij}}

\newcommand{\mij}{m_{V,ij}}
\newcommand{\Rij}{R_{0,ij}}
\newcommand{\half}{{\textstyle{\frac{1}{2}}}}

\vspace{2cm}

\title{
\bf Perturbative and non-perturbative studies of the SU(2)-Higgs model  \\
on lattices with asymmetric lattice spacings}
\author{%
F.~Csikor \\
\institute{Institute for Theoretical Physics, E\"otv\"os University,\\
H-1088 Budapest, Hungary }\\  
Z.~Fodor 
\thanks{On leave from Institute for Theoretical Physics,
E\"otv\"os University, H-1088 Budapest, Hungary}\\
\institute{KEK, Theory Group, 1-1 Oho, Tsukuba 305, \\ Japan}\\
J.~Heitger 
\thanks{Present address: DESY Zeuthen, Platanenallee 6, D-15738 
Zeuthen, Germany }\\
\institute{Institut f\"ur Theoretische Physik I, Universit\"at M\"unster,\\
D--48149 M\"unster, Germany
}}
\date{}
\begin{document}
\maketitle

\vspace{3.cm}
\begin{abstract} 
We present a calculation of the ${\cal O}(g^2,\lambda)$ perturbative 
corrections to the
coupling anisotropies of the SU(2)-Higgs model on lattices
with asymmetric lattice spacings.
These corrections are obtained by a one-loop calculation
requiring the rotational invariance of the
gauge and Higgs boson propagators in the continuum limit.
The coupling anisotropies are also determined from numerical 
simulations of the model on appropriate lattices. The one-loop perturbation 
theory and the simulation results agree with high accuracy. 
It is demonstrated that rotational invariance is also restored
for the static potential determined from 
space--space and space--time Wilson  loops.

PACS Numbers: 11.15.Ha, 12.15.-y\\
\end{abstract}
\vfill

\section{Introduction}

At high temperatures the electroweak symmetry is restored.
Since the baryon violating processes are unsuppressed at high temperatures,
the observed baryon asymmetry of the universe has finally been
determined at the electroweak phase transition \cite{kuzmin}.

In recent years quantitative studies of the electroweak phase
transition have been carried out by means of resummed
perturbation theory
and lattice Monte Carlo simulations \cite{arnold}--\cite{CFHHJM97}.
In the SU(2)-Higgs model for Higgs masses ($m_H$) below 50 GeV,
the phase transition is predicted by the perturbation theory to be
of first order. However, it is difficult to give a  definite perturbative 
statement 
for physically more interesting masses, e.g.
$m_H>80$ GeV. Due to the bad infrared properties of the theory,
the perturbative approach breaks down in
this parameter region. A systematic and fully
controllable treatment is necessary, which can be achieved by
lattice simulations.

For smaller Higgs boson masses ($m_H<50$ GeV)
the phase  transition is quite
strong and relatively easy to study on the lattice.
For larger $m_H$ (e.g. $m_H=80$ GeV)
the phase transition gets weaker, the lowest excitations have
masses small compared to the temperature, $T$. From this
feature one expects that a finite temperature simulation
on an isotropic lattice would
need several hundred lattice points in the spatial directions
even for $L_t=2$ temporal extension. These kinds of lattice sizes
are out of the scope of the present numerical resources.

One possibility to solve the problem of these different scales is to
integrate out the heavy, ${\cal O}(T)$ modes perturbatively,
and analyse the obtained theory on the lattice. This strategy
turned out to be quite successful, and both its perturbative
and lattice features have been studied by
several groups \cite{kajantie}--\cite{philipsen}.
Even more: \cite{shap} predicts that somewhere above  80 GeV Higgs mass 
the first order phase transition does not take place any further, 
the two phases can be continuously connected. Ref.~\cite{Leipzig} gives 
estimates of 
the end-point of the phase transition in the framework of the reduced 3-d 
approach.

With this paper we follow another approach (analytic
and Monte Carlo) to handle this
two-scale problem. We will use the simple idea that
finite temperature field theory can be
conveniently studied on asymmetric lattices, i.e. lattices
with different spacings in temporal ($a_t$) and spatial
($a_s$) directions.  This method solves the two-scale
problem in a natural way \cite{bender}. Another advantage is,
well-known and used in QCD, that this formulation
makes an independent variation of the temperature ($T$) and
volume ($V$) possible. The perturbative corrections
to the coupling anisotropies are known in QCD (see
refs. \cite{karsch82,karsch89}). Performing a similar analysis for the 
SU(2)-Higgs model, we have presented in our earlier paper \cite{Cs-F} 
the perturbative corrections to the coupling anisotropies for this case, too.
 Here we give  
details of the perturbative calculation of \cite{Cs-F} and 
by numerical simulation on lattices with anisotropic lattice spacings 
calculate the  coupling asymmetries for the practically 
reasonable parameters of the SU(2)-Higgs model.
As we will show at $m_H \approx 80$ GeV the one-loop perturbative and 
the non--perturbative coupling asymmetries agree very well.

There is an essential difference between pure gauge theories and 
the SU(2)-Higgs model. In the former case any value (within a certain range) 
of the space and time coupling constant ratio does correspond to a 
meaningful theory, the actual value of the ratio corresponds to a 
definite value of the ratio of the space and time lattice spacings.
On the other hand, in case of the SU(2)-Higgs model for a fixed value of the 
space and time gauge coupling ratio (determining the  ratio of the space 
and time lattice spacings) one has to fix the ratio of the 
space and time hopping parameters to a definite value, in order 
to ensure that the 
theory makes sense. A convenient  way to do this is to require that the 
ratio of space and time gauge boson masses should be equal to the ratio of 
 space and time Higgs boson masses. Such a choice of the parameters is 
a precondition to both the perturbative calculations and the numerical 
simulations.

The plan of this paper  is as follows. Section II. deals with the perturbative 
analysis. In subsection II.A we give the lattice
action of the model on asymmetric lattices and discuss
perturbation theory in the anisotropic lattice case.
Subsection II.B contains the calculation of the critical hopping parameter 
and of the
wave function quantum correction terms, which give
the quantum corrections to the anisotropy 
parameters. In subsection II.C a discussion of the finite temperature continuum 
limit is  given. The optimal choice of the ratio of space and time lattice 
spacings is determined by perturbative techniques.  
Section III. contains our non-perturbative analysis. 
 Subsection III.A gives the basic points of our MC simulations. Subsection 
III.B  
deals with the mass determinations from the correlation functions.
In subsection III.C we present the results on Wilson loop simulations and 
the static potential. In subsection III.D we finally evaluate the 
non-perturbative 
asymmetries and compare them with the perturbative results.
Section IV. is a summary   and outlook.

\section{Lattice action and perturbation theory}

In this section we discuss lattice perturbation theory. Since we did not 
find the Feynman rules for the anisotropic lattice spacing case in the 
literature, we present some details of lattice perturbation theory. 
After that we determine the critical hopping parameter and the anisotropies in 
one-loop perturbation theory. The continuum limit and the optimal choice of 
the ratio of space- and time-like lattice spacings is also discussed.

\subsection{Action in continuum notation, gauge fixing, propagators}

For simplicity, we use equal lattice spacings in the three
spatial directions ($a_i=a_s,\ i=1,2,3$) and another spacing
in the temporal direction ($a_4=a_t$). The asymmetry of the
lattice spacings is characterized by the asymmetry factor $\xi=a_s/a_t$.
The different lattice spacings can be ensured by
different coupling strengths in the action for time-like and space-like
directions. The action reads
\begin{eqnarray}\label{lattice_action}
S[U,\varphi] &=& \beta_s \sum_{sp}
\left( 1 - {1 \over 2} {\rm Tr\,} U_{sp} \right)
+\beta_t \sum_{tp}
\left( 1 - {1 \over 2} {\rm Tr\,} U_{tp} \right)
\nonumber \\
&&+ \sum_{x\in\Lambda} \left\{ {1 \over 2}{\rm Tr\,}(\varphi_x^+\varphi_x)+
\lambda \left[ {1 \over 2}{\rm Tr\,}(\varphi_x^+\varphi_x) - 1 \right]^2
\right. \nonumber \\
&&\left.
-\kappa_s\sum_{\mu=1}^3
{\rm Tr\,}(\varphi^+_{x+\hat{\mu}}U_{x,\mu}\,\varphi_x)
-\kappa_t {\rm Tr\,}(\varphi^+_{x+\hat{4}}U_{x,4}\,\varphi_x)\right\},
\end{eqnarray}
where  $\Lambda$ stands for the lattice points, $U_{x,\mu}$ denotes the SU(2) 
gauge link variable,  $U_{sp}$ and
$U_{tp}$
the path-ordered product of the four $U_{x,\mu}$ around a
space-space and space-time plaquette, respectively.
The symbol $\varphi_x$ stands for the Higgs field.

The values of anisotropies defined as 
\begin{equation}\label{uj}
\gamma_\beta^2={\beta_t \over \beta_s}\ \ \ , \ \ \ \
\gamma_\kappa^2={\kappa_t \over \kappa_s}
\end{equation}
are choosen to correspond to given values of the asymmetry $\xi$.
In perturbation theory this can be ensured order by order in the 
loop expansion, 
requiring that in the limit $a_s , \, a_t \rightarrow 0$ with the ratio 
$\xi=a_s/a_t$ fixed, certain physical quantities show rotation symmetry 
on submanifolds of the bare coupling space satisfying 
$\gamma_\beta =const$, $\gamma_\kappa =const$. This 
procedure leads to a formal double expansion in $g^2$ and $\lambda$ of the 
anisotropies:
\begin{equation}\label{tree}
\gamma_\beta^2=
\xi^2\left[1+c_\beta(\xi)g^2+b_\beta(\xi)\lambda
+{\cal O}(g^4,\lambda^2)\right],\ \
\gamma_\kappa^2=
\xi^2\left[1+c_\kappa(\xi)g^2+b_\kappa(\xi)\lambda
+{\cal O}(g^4,\lambda^2)\right].
\end{equation}
Here $g$ is the bare gauge coupling of the theory with symmetric lattice 
spacings in standard notation. 
(Note that $c_\beta(1)=c_\kappa(1)=b_\beta(1)=b_\kappa(1)=0$.) 
In this double expansion we use the formal power counting
$\lambda \sim g^2$.
In general, fixing  $\gamma_\beta(\xi)$ and
$\gamma_\kappa(\xi)$ to ensure rotation symmetry 
should be done non-perturbatively. In the non-perturbative framework the 
definition of 
$\xi $ is given as the ratio of space and time direction lattice unit 
correlation lengths. This non-perturbative analysis 
will be the topic of section III., where we choose the values of the 
bare copuling ratios (\ref{uj}) to ensure 
that the Higgs  and gauge boson correlation lengths in physical units
be  the same in the different directions.
This idea can be applied in perturbation theory as well (see e.g.
\cite{karsch89}), and we will follow this method in our analysis, 
too. 

Elaborating perturbation theory we follow the usual steps (see e.g. 
\cite{montvay}, \cite{MMu} for the isotropic SU(2)-Higgs model). The 
only complication is 
that we have to keep track of the different lattice spacings and couplings. \\
First we consider the gauge part of the action.
We will use the same notation as applied by the calculation 
\cite{karsch82} for the pure gauge theory,

\begin{equation}\label{link}
U_{x,\mu}=\exp\left(ia_\mu g_\mu \frac{\tau_r}{2}A_\mu^r (x)\right),
\end{equation}
where $r$ is summed over 1,2,3, while $\mu$ is not summed, moreover $a_\mu=a_s$,
$g_\mu=g_s$ for $\mu=1,2,3$ and $a_4=a_t$, $g_4=g_t$. We have also 
\begin{equation}\label{beta}
\beta_s=\frac{4}{\xi}\frac{1}{g_s(\xi)^2},  \ \ 
\beta_t=4 \xi \frac{1}{g_t(\xi)^2}
\end{equation}
as the connection to the lattice parameters.
The expansions for $g_s$, $g_t$ read:
\begin{equation}\label{g_s}
g_s(\xi)^2=g^2 (1-c_s (\xi) g^2 -b_s (\xi) \lambda +
{\cal O}(g^4 ,\lambda^2 )),
\end{equation}
\begin{equation}\label{g_t}
g_t(\xi)^2=g^2 (1+c_t (\xi) g^2 +b_t (\xi) \lambda +
{\cal O}(g^4 ,\lambda^2 )),
\end{equation}
where $g^2=g_t^2 (\xi=1)=g_s^2 (\xi=1)$ and 
$c_\beta (\xi)=c_t (\xi)-c_s (\xi)$, 
$b_\beta (\xi)=b_t (\xi)-b_s (\xi)$.

We write 
\begin{equation}\label{small_a}
U_{x,\mu}=a_\mu^0 (x)+i\tau_r a_\mu^r (x),
\end{equation}
where 
\begin{equation}\label{reszl}
a_\mu^0 (x)=\cos\left(\frac{a_\mu g_\mu |A_\mu (x) |}{2}\right), \ \ \ 
a_\mu^r (x)=\frac{A_\mu^r (x)}{|A_\mu (x)|}\sin\left(
\frac{a_\mu g_\mu |A_\mu (x) |}{2}\right),
\end{equation}
with $|A_\mu (x)|=\sqrt {A_\mu^r (x)A_\mu^r (x)}$.
The expansion is given by:
\begin{equation}\label{exp1}
a_\mu^0 (x)=1-\frac{(a_\mu g_\mu )^2}{8}A_\mu^r A_\mu^r + 
\frac{(a_\mu g_\mu )^4}{384} A_\mu^r A_\mu^r A_\mu^s A_\mu^s +{\cal O}(g^6),
\end{equation}
\begin{equation}\label{exp2}
a_\mu^r (x)=\frac{a_\mu g_\mu}{2}A_\mu^r - \frac{(a_\mu g_\mu )^3}{48}
A_\mu^r A_\mu^s A_\mu^s +{\cal O}(g^5).
\end{equation}

Inserting eq. (\ref{small_a}) into the plaquette parts of the lattice action 
we get the parts of the action containing odd numbers of gauge boson fields   
($S_{pl}^{odd}$) and even numbers of gauge boson fields   ($S_{pl}^{even}$).
They read:
\begin{eqnarray}\label{odd}
S_{pl}^{odd}=- \sum_{x\in\Lambda} \sum_{\mu < \nu} \beta_{\mu \nu} 
\left\{ \epsilon_{prs} (a_\nu^0 (x+\hat \mu a_\mu )a_\mu^p (x)+
a_\mu^0 (x)a_\nu^p (x+\hat \mu a_\mu))a_\mu^r (x+\hat \nu a_\nu)a_\nu^s (x)
\right. \nonumber\\
\left. -\epsilon_{prs} a_\mu^p (x)a_\nu^r (x+\hat \mu a_\mu)
(a_\nu^0 (x) a_\mu^s (x+\hat \nu a_\nu)+
a_\mu^0 (x+\hat \nu a_\nu )a_\nu^s (x)) \right\},
\end{eqnarray}
\begin{eqnarray}
S_{pl}^{even}=- \sum_{x\in\Lambda} \sum_{\mu < \nu} \beta_{\mu \nu}
\left\{ a_\mu^0 (x)a_\nu^0 (x+\hat \mu a_\mu )a_\mu^0 (x+\hat \nu a_\nu ) 
a_\nu^0 (x)
-a_\mu^0 (x)a_\nu^0 (x+\hat \mu a_\mu )a_\mu^r (x+\hat \nu a_\nu ) 
a_\nu^r (x) \right. \nonumber\\
\left. -a_\mu^0 (x+\hat \nu a_\nu ) a_\nu^0 (x)
a_\mu^r (x) a_\nu^r (x+\hat \mu a_\mu )
+a_\mu^r (x) a_\nu^r (x+\hat \mu a_\mu )
a_\nu^s (x+\hat \nu a_\nu )a_\nu^s (x) \right. \nonumber\\
\left. +(a_\nu^0 (x+\hat \mu a_\mu )a_\mu^r (x)+a_\mu^0 (x) 
a_\nu^r (x+\hat \mu a_\mu ))
(a_\nu^0 (x) a_\mu^r (x+\hat \nu a_\nu )+a_\mu^0 (x+\hat \nu a_\nu )
a_\nu^r (x)) \right. \nonumber \\
\left. +(-\delta_{ps} \delta_{rt} + \delta_{pt} \delta_{rs} )
a_\mu^p (x) a_\nu^r (x+\hat \mu a_\mu ) a_\mu^s (x+\hat \nu a_\nu ) a_\nu^t (x)
\right\}, \nonumber
\end{eqnarray}
\begin{eqnarray}\label{even}
\end{eqnarray}
where p,r,s,t=1,$\ldots,3$  
and $\beta_{\mu \nu}$ is equal to $\beta_s $ for space indices and equal to 
$\beta_t $ for one space and one time index. 

 The integration measure for the 
gauge variables also contributes to the action.
\begin{eqnarray}\label{measure}
d^3 U_{x,\mu} = \frac{1}{\pi^2}d^4 a_\mu^S (x) \delta 
(a_\mu^T (x) a_\mu^T (x) -1) \nonumber \\
\rightarrow d^3 A_\mu^r (x) \exp \left(\log \left(
\frac{\sin^2 (\frac{g_\mu a_\mu}{2}
| A_\mu |)}{\frac{1}{4}g_\mu^2 a_\mu^2 | A_\mu |^2}\right)\right),
\end{eqnarray}
where capital letters run from 0 to 3 and a sum over T=0,$\ldots,3$ is 
understood. The contribution to the action reads:
\begin{eqnarray}\label{measure1}
S_m =- \sum_{x\in\Lambda} \sum_{\mu=1}^4 
\log \left(\frac{\sin^2 \left(\frac{g_\mu a_\mu}{2} 
| A_\mu |\right)}{\frac{1}{4}g_\mu^2 a_\mu^2 | A_\mu |^2}  \right)
= \sum_{x\in\Lambda} \sum_{\mu=1}^4
\frac{g_\mu^2 a_\mu^2}{12} A_\mu^r (x)A_\mu^r (x)+{\cal O}(g^4).
\end{eqnarray} 
Next we consider the pure scalar part of the action. 
Introducing the notation
\begin{equation}
\varphi _x =H_0(x)+i\tau_r \pi_r (x),
\end{equation}
it  reads:
\begin{eqnarray}\label{SHiggs}
S_H =\sum_x \left\{ H_0 (x)^2+\pi_r (x)\pi_r (x)
+\lambda (H_0 (x)^2+
\pi_r (x)\pi_r (x)-1)^2 \right. \nonumber \\ \left.
-2\kappa_s \sum_{\mu=1}^3 
\left(H_0 (x+\hat \mu a_\mu )H_0 (x)-\pi_r (x+\hat \mu a_\mu)\pi_r (x)\right)  
\right. \nonumber \\ \left.
-2 \kappa_t (H_0 (x+\hat 4 a_4 ) H_0 (x)-\pi_r (x+\hat 4 a_4)\pi_r (x))
\right\}.
\end{eqnarray}
Assuming that the $H_0$ field has a non-zero vacuum expectation value $v$, we 
write:
\begin{equation}
H_0 (x)=H(x)+v.
\end{equation}
Moreover we introduce the notations
\begin{equation}
g_0=6\lambda\frac{1}{\kappa_s^2}\frac{a_t}{a_s}, \ \ \ 
\lambda_c =\frac{g_0}{24}
\end{equation}
and the continuum fields
\begin{eqnarray}
H_c (x)=\left(\frac{2\kappa_s}{a_s a_t}\right)^{\frac{1}{2}} H(x), \ \ \ 
\pi_c^r (x)=\left(\frac{2\kappa_s}{a_s a_t}\right)^{\frac{1}{2}} 
\pi^r (x), \ \ \
v_c =\left(\frac{2\kappa_s}{a_s a_t}\right)^{\frac{1}{2}} v.
\end{eqnarray}
Using these we find the scalar part of the action using continuum 
variables to be
\begin{eqnarray}\label{SH_cont}
S_H = a_s^3 a_t \sum_{x\in\Lambda} \frac{1}{2}\left\{ \sum_{i=1}^3 
(\nabla_i H_c (x) \nabla_i H_c (x)+\nabla_i \pi_c^r (x) \nabla_i \pi_c^r (x))\right.
\nonumber \\\left.
+\frac{\gamma_\kappa^2 }{\xi^2}
(\nabla_4 H_c (x)\nabla_4 H_c (x)+\nabla_4 \pi_c^r (x) \nabla_4 \pi_c^r (x))
\right. \nonumber \\\left.
+m_0^2 ((H_c(x) +v_c )^2 +\pi_c^r (x)\pi_c^r (x)) 
+\frac{g_0}{12}((H_c (x)+v_c )^2 +\pi_c^r (x)\pi_c^r (x))^2\right\},
\end{eqnarray}
where
\begin{equation}
a_s^2 m_0^2 =\frac{1-2\lambda }{\kappa_s} -6-2\gamma_\kappa^2,
\end{equation}
and
\begin{equation}
\nabla_\mu f(x)=\frac{f(x+\hat \mu a_\mu )-f(x)}{a_\mu}
\end{equation}
is the lattice derivative.

Putting $v_c =0$ above corresponds to the symmetric phase, 
in this case  $m_0^2 > 0$. 
Determining  $v_c $ from the non-trivial minimum of the scalar potential 
one gets
\begin{equation}
v_c^2=-\frac{6 m_0^2}{g_0} \ \ \ {\rm for} \ \ \ m_0^2 < 0.
\end{equation}
Introducing $m_{H,0}^2 =-2 m_0^2$ we obtain finally:
\begin{eqnarray}\label{SH_cont_fin}
S_H = a_s^3 a_t \sum_{x\in\Lambda} \frac{1}{2}\left\{ \sum_{i=1}^3
(\nabla_i H_c (x) \nabla_i H_c (x)+\nabla_i \pi_c^r (x) \nabla_i \pi_c^r (x))
\right.
\nonumber \\\left.
+\frac{\gamma_\kappa^2 }{\xi^2}
(\nabla_4 H_c (x)\nabla_4 H_c (x)+\nabla_4 \pi_c^r (x) \nabla_4 \pi_c^r (x))
\right. \nonumber \\\left.
+m_{H,0}^2 H_c(x)^2 
+\frac{g_0}{12}(H_c (x)^2 +\pi_c^r (x)\pi_c^r (x))^2
+\frac{g_0 v_c}{3} H_{c,0} (x) ( H_{c,0}^2 (x)+\pi_c^r (x)\pi_c^r (x))
\right \}.
\end{eqnarray}

Now we consider the  gauge--scalar interaction:              
\begin{eqnarray}\label{gauge-H}
S_i = \sum_{x\in\Lambda} \left \{ -\kappa_s \sum_{i=1}^3 {\rm Tr\,}(
\varphi _{x+\hat i a_i}^+ (U_{x,i}-1)\varphi _x )-
\kappa_t {\rm Tr\,}( \varphi _{x+\hat 4 a_t }^+ 
(U_{x,4}-1)\varphi _x ) \right \}.
\end{eqnarray}
Introducing  continuum variables  we obtain
\begin{eqnarray}\label{gauge-H_c}
S_i =a_s^3 a_t \sum_{x\in\Lambda} \left \{
\sum_{i=1}^3 (a_i^0 (x) -1)\left(-\frac{1}{a_s^2} \left(v_c^2+2v_c H_c(x)+H_c(x) H_c(x)+
\pi_c^r (x) \pi_c^r (x)\right)
\right. \right. \nonumber\\ \left. \left.
-\frac{v_c}{a_s}\nabla_i H_c (x)-\frac{1}{a_s} 
\left(H_c (x) \nabla_i H_c (x) +\pi_c^r (x) \nabla_i \pi_c^r (x) \right)\right) 
\right. \nonumber\\ \left.
+\frac{\gamma_\kappa^2 }{\xi^2 }(a_4^0 (x)-1)\left(-\frac{1}{a_t^2} 
\left(v_c^2+2v_c H_c(x)+
H_c(x) H_c(x)+\pi_c^r (x) \pi_c^r (x)\right)-\frac{v_c}{a_t}\nabla_4 H_c (x)-
\right. \right.\nonumber\\ \left. \left. 
\frac{1}{a_s}\left(H_c (x) \nabla_4 H_c (x) +\pi_c^r (x) \nabla_4\pi_c^r (x) 
\right) \right) 
\right. \nonumber\\ \left.
+\frac{1}{a_s}\sum_{i=1}^3 a_i^r (x) \left(\epsilon_{rst} \pi_c^s (x) \nabla_i 
\pi_c^t (x) +\pi_c^r (x) \nabla_i H_c (x) -(H_c (x) +v_c) \nabla_i \pi_c^r (x)
\right)
\right. \nonumber \\  \left.
+\frac{a_t }{a_s^2}\gamma_\kappa^2  a_4^r (x) \left(\epsilon_{rst} \pi_c^s (x) 
\nabla_4 \pi_c^t (x) +\pi_c^r (x) \nabla_4 H_c (x) -(H_c (x) +v_c) 
\nabla_4 \pi_c^r (x)\right) \right\}.
\end{eqnarray}
In perturbation theory the gauge has to be fixed. We use as  the gauge fixing function 
\begin{equation}\label{GFfun}
f_r (x)=\sum_{i=1}^3 \frac{a_i^r (x)-a_i^r (x-\hat i a_i )}{a_i^2}+
\frac{\gamma_\kappa^2 (a_4^r (x)-a_4^r (x-\hat 4 a_t ))}{a_t^2}+
\frac{\alpha v_c g^2 \pi_c^r }{4} ,
\end{equation}
which is a lattice version of the well known continuum $R_\xi$ gauge fixing 
function. In eq. (\ref{GFfun}) $\alpha$ is the gauge parameter. 
This choice ensures that the mixed second order term 
in $A_\mu^r (x)$ and $\pi_c^r (x) $ will drop out
from the sum of the gauge--scalar interaction and the gauge fixing parts of 
the action. 
We obtain
\begin{equation}\label{GF}
S_{gf} =-\frac{2}{\alpha g^2} a_s^3 a_t \sum_{x\in\Lambda} f_r (x) f_r (x),
\end{equation}
\begin{eqnarray}\label{FP}
S_{FP} = a_s^3 a_t \sum_{x\in\Lambda} 
\sum_{i=1}^3 \left \{ \frac{1}{a_s^2} 
\left \{ (\bar c_r (x) -\bar c_r (x+\hat i a_i))a_i^0 (x) 
(c_r (x) + c_r (x+ \hat i a_i))
\right. \right. \nonumber \\ \left. \left.
+\epsilon_{rst}
(\bar c_r (x) +\bar c_r (x+\hat i a_i))a_i^s (x) 
(c_t (x) - c_t (x+ \hat i a_i))\right \}
\right. \nonumber \\ \left.
+ \frac{\gamma_\kappa^2 }{\xi^2 a_t^2}
\left \{(\bar c_r (x) -\bar c_r (x+\hat 4 a_t))a_4^0 (x)
(c_r (x) + c_r (x+ \hat 4 a_t))
\right. \right. \nonumber \\ \left. \left.
+\epsilon_{rst}
(\bar c_r (x) +\bar c_r (x+\hat 4 a_t))a_4^s (x)
(c_t (x) - c_t (x+ \hat 4 a_4))\right \}
\right. \nonumber \\ \left.
-\alpha \frac{g^2 v_c }{4} \left \{ \bar c_r (x) c_r (x) 
(H_c (x) +v_c )+\epsilon_{rst}
\bar c_r (x) c_s (x) \pi_c^t (x) \right \} \right \}.
\end{eqnarray}

The final form of the continuum notation action reads:
\begin{equation}\label{action_fin}
S_{cont} =S_{pl}^{odd} + S_{pl}^{even} + S_m + S_H + S_i + S_{gf} + S_{FP} ,
\end{equation}
where the individual terms are given in eqs. (\ref{odd}, \ref{even}, 
\ref{measure1}, \ref{SH_cont_fin},
\ref{gauge-H_c}, \ref{GF}, \ref{FP}). 
The vertices of perturbation theory may be easily obtained from 
eq. (\ref{action_fin}).  As usual in lattice perturbation theory 
we have  new vertices proportional to $g^n$ for $n \ge 2$. 

The formulae to compute the Fourier transforms are as follows.
For the gauge field:
\begin{equation}\label{FT1}
\tilde a_{k,\mu}^R =a_s^3 a_t \sum_{x\in\Lambda} \exp 
\left(-i(k,x)-\frac{ia_\mu}{2}k_\mu \right)
a_\mu^R (x),
\end{equation}
where $(k,x)=2\pi (\frac{\nu_1 l_1}{L_1} + \cdots +\frac{\nu_4 l_4}{L_4} )$, 
$R=0,1,2,3$  and  
\begin{equation}
k_\mu = \frac{2\pi}{L_\mu a_\mu } \nu_\mu , \ \ 
x_\mu =a_\mu l_\mu,
\end{equation}
moreover the integers $\nu_\mu$,  $ (l_\mu)$ take values from $0,1, 
\dots ,L_\mu -1 $. 

The inverse relation is:
\begin{equation}\label{FT2}
a_\mu^R (x)=\frac{1}{L_1 L_2 L_3 L_4}\frac{1}{a_1 a_2 a_3 a_4 } 
\sum_{\nu_\mu }\exp \left(i(k,x)+\frac{ia_\mu}{2}k_\mu \right) 
\tilde a_{k,\mu}^R.
\end{equation}
For a lattice infinite in all directions we have 
\begin{equation}\label{FT3}
a_\mu^R (x) =\frac{1}{(2\pi )^4 }\prod_{\rho =1}^4 \int_{
-\pi/a_\rho}^{\pi/a_\rho} 
dp_\rho  \exp\left( i p\cdot x+ ip_\mu a_\mu \right) \tilde a_{k,\mu}^R ,
\end{equation}
and $p\cdot x =\sum_{\nu =1}^4 p_\nu x_\nu $. 

For scalar fields we have similar formulae, however, the second terms 
are missing in the exponents of eqs. (\ref{FT1},\ref{FT2},\ref{FT3}).

Our aim is to perform perturbative calculations. The first step is to 
write down the propagators of the fields from parts quadratic in the fields 
of the action. We want to determine the tree-level propagators, 
which are zeroth order in $g$ and $\lambda$. Since $\gamma_\beta $ and 
$\gamma_\kappa $ do depend on the couplings, we use $\xi$ 
in the propagators as their tree-level values. 
 The remaining correction terms from  $\gamma_\beta $ and $\gamma_\kappa $ 
are quadratic in the fields and 
 give two particle vertices similarly to the measure term in the action 
in the isotropic case. These will be absorbed by the kinetic parts of the 
propagators (see later in eqs. (\ref{Higgs_2p})--(\ref{W3-propagator})).

The inverse tree-level propagators in momentum space have the following 
forms.\\      
For the Higgs boson
\begin{equation}\label{H-propagator}
\tilde\Delta_{H,0}(p)^{-1}=\sum_{i=1}^4 {\hat p}_i^2+m_{H,0}^2,
\end{equation}
for the Goldstone bosons
\begin{equation}\label{Goldstone-propagator}
\tilde\Delta_{\pi_r^c,0}(p)^{-1}=\sum_{i=1}^4 {\hat p}_i^2
-\alpha m_{W,0}^2,
\end{equation}
for the gauge boson
\begin{equation}\label{W-propagator}
\tilde\Delta_{W,0,\mu\nu}^{ab}(p)^{-1}
=\delta^{ab}\delta_{\mu\nu}\left[
m_{W,0}^2+\sum_{i=1}^4 {\hat p}_i^2 \right]- {\hat p}_\mu {\hat p}_\nu 
\frac{1+\alpha}{\alpha},
\end{equation}
for the ghost
\begin{equation}\label{FP-propagator}
\tilde\Delta_{FP,0}(p)^{-1}=\sum_{i=1}^4 {\hat p}_i^2
-\alpha m_{W,0}^2,                   
\end{equation}
where
\begin{equation}
{\hat p}_i={2 \over a_s}\sin{a_sp_i \over 2},\ \ \
{\hat p}_4={2 \over a_t}\sin{a_tp_4 \over 2}.
\end{equation}
The masses have the following expressions in terms 
of other parameters:
\begin{equation}
m_{H,0}^2=-{2 \over a_s^2}
\left[{1-2\lambda \over \kappa}\xi-6-2\xi^2\right],\ \ \
m_{W,0}^2={m_{H,0}^2\kappa^2 \over 2\lambda\xi\beta}
=\frac{v_c^2 g^2}{4}.
\end{equation}

\subsection{Critical hopping parameter and anisotropy parameters}

The main goal of the paper is to perform a ${\cal O} (g^2 , \lambda )$
 analysis of the
theory defined by eq. (\ref{lattice_action}).
This means first the determination of
the mass-counterterms. One wants to tune the bare parameters in a way 
to ensure that
the one-loop renormalized masses are finite in the continuum limit (however,
their values in lattice units do vanish, $a_s m_{ren}=0$ for
$a_s \rightarrow 0, \ a_t \rightarrow 0, \ \xi =$ fixed). At the same time the 
vacuum expectation value of the
scalar field will be also zero in lattice units
($a_s v_c=0$), i.e. we are at  the phase transition  point 
between the spontaneously broken Higgs phase and the SU(2) symmetric phase.
The condition is fulfilled by an appropriate choice of the hopping parameter
(critical hopping parameter). The ratios of the couplings ($\gamma_\beta$
and $\gamma_\kappa$) are still free parameters and can be fixed
by two additional conditions. We demand rotational (Lorenz)
invariance for the scalar and gauge boson propagators on the
one-loop level. This ensures that the propagators with one-loop
corrections have the same form in the $z$-- and $t$-- directions.
Clearly, arbitrary couplings for different directions
in eq. (\ref{lattice_action}) would not lead to such rotationally
invariant two-point functions.

\begin{figure}
\begin{center}
\epsfig{file=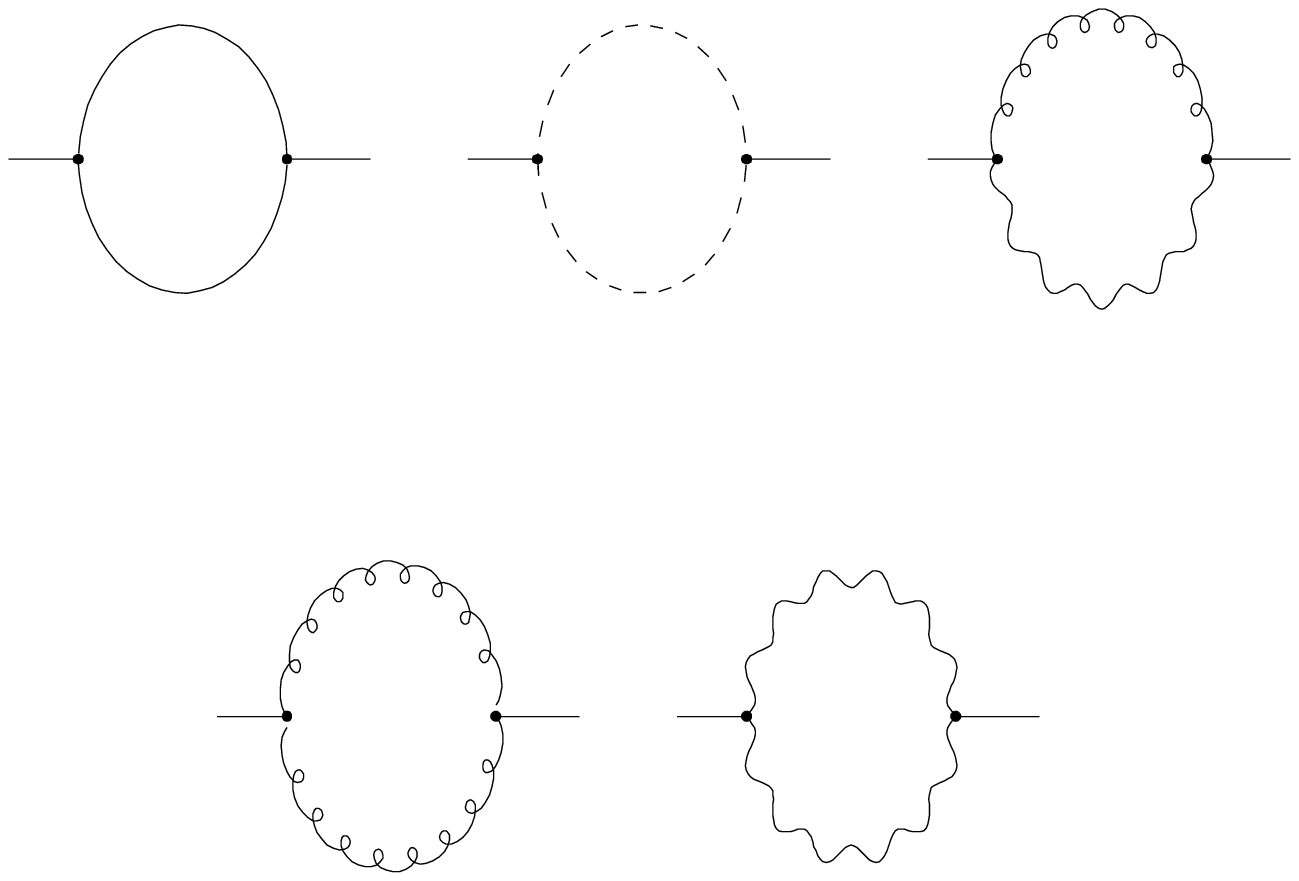,width=6.4cm}
\epsfig{file=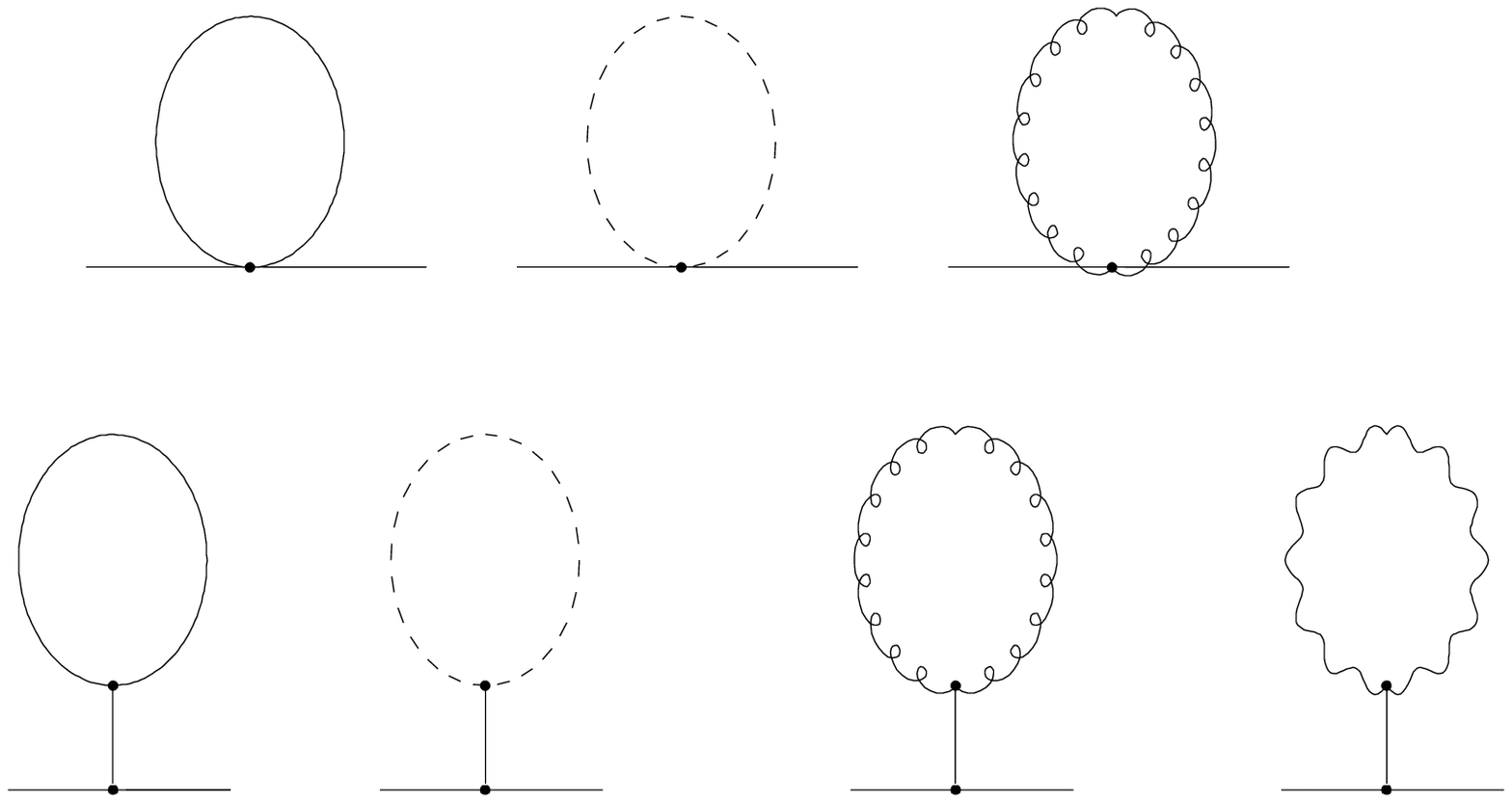,width=8cm}
\caption{\label{Higgs_self}
{\sl Higgs boson self energy graphs. Solid lines stand for Higgs, dashed
lines for Goldstone, curly lines for vector bosons and wavy lines denote ghosts.
}}
\end{center}\end{figure}

The most straightforward method to determine the  transition point is
the use of the effective potential. 
 The condition $d^2V_{eff}(\Phi=0)/d\Phi^2=0$ gives a simple, gauge invariant  
expression for the value of the critical hopping parameter, which is 
exact in the continuum limit. 
The relevant formulae are given in Landau gauge in \cite{Cs-F}.
Here we present the general $R_\xi$ effective potential:

\begin{eqnarray}
V_{eff}(\Phi)={m_0^2 \over 2}\Phi^2+\lambda_c\Phi^4
+\int_k \left[{1 \over 2} \log({\hat k}^2+m_0^2+12\lambda_c\Phi^2)
\right. \nonumber \\ \left.
+{24 \lambda_c\over {16 \lambda_c-\alpha g^2}} 
\log ({\hat k}^2+m_0^2+4\lambda_c\Phi^2-\alpha g^2\Phi^2/4)
\right. \nonumber \\ \left.
+6 \log({\hat k}^2+g^2\Phi^2/4) +
{3 \over 2}\log\left
(\frac{{\hat k}^2-\alpha g^2\Phi^2/4}{{\hat k}^2+g^2\Phi^2/4}\right)
\right. \nonumber \\ \left.
-{3 \over 2} \log({\hat k}^2-\alpha g^2\Phi^2/4)\right],
\end{eqnarray}
where
\begin{equation}
\int_k \equiv {1 \over (2\pi)^4}
\int_{-\pi/a_s}^{\pi/a_s}d^3k\int_{-\pi/a_t}^{\pi/a_t}dk_4.
\end{equation}
Alternatively, one may calculate the one-loop corrections to the masses 
and require that the renormalized masses be zero in lattice units in the 
limit of  zero lattice spacing and fixed $\xi =a_s /a_t$, as explained 
above. 

First we consider the corrections arising from the 
 two-point interaction vertices. In addition to these there are the one-loop 
corrections, which we evaluate later on. 
Including           the two-point interaction vertex corrections the momentum
squared sums in 
eqs.  (\ref{H-propagator}), (\ref{Goldstone-propagator}), 
(\ref{FP-propagator})
modify to
\begin{equation}\label{Higgs_2p}
\sum_{i=1}^4 {\hat p}_i^2 \longrightarrow
\sum_{i=1}^3 {\hat p}_i^2 +\frac{\gamma_\kappa^2}{\xi^2} {\hat p}_4^2
\end{equation}
for the Higgs, Goldstone and ghost propagators.
The gauge boson inverse propagator becomes more complicated:
\begin{equation}\label{W1-propagator}
\tilde\Delta_{W,ij}^{ab}(p)^{-1}
=\delta^{ab} \left[ \delta_{ij}
\left(\frac{v_c^2 g_s^2}{4}+\sum_{i=1}^3 {\hat p}_i^2
 +{\gamma_\beta^2 \over \xi^2}{\hat p}_4^2 \right)
-{\hat p}_i {\hat p}_j \left(1+\frac{g_s^2}{\alpha g^2} \right) \right],
\end{equation}

\begin{equation}\label{W2-propagator}
\tilde\Delta_{W,i4}^{ab}(p)^{-1}=\tilde\Delta_{W,4i}^{ab}(p)^{-1}
=-\delta^{ab}{\hat p}_i {\hat p}_4 \left[\frac{\gamma_\beta}{\xi} +
\frac{g_s g_t}{\alpha g^2} \frac{\gamma_\kappa^2}{\xi^2 } \right],
\end{equation}

\begin{equation}\label{W3-propagator}
\tilde\Delta_{W,44}^{ab}(p)^{-1}=\delta^{ab}\left[
\frac{v_c^2 g_t^2}{4}\frac{\gamma_\kappa^2}{\xi^2 }+\sum_{i=1}^3 {\hat p}_i^2
-{\hat p}_4^2\frac{g_t^2}{\alpha g^2} \frac{\gamma_\kappa^4}{\xi^4 } \right].
\end{equation}

Let us now consider the self-energy corrections to the Higgs mass. 
The relevant diagrams are shown in figure 1.
Evaluating all graphs we obtain at zero Higgs four-momentum, independent 
of the gauge choosen (i.e. independent on $\alpha$):
\begin{equation}\label{Higgs-mass}
a_s^2 (m_H^R )^2 =a_s^2 m_H^2 -\left(2g_0+\frac{9}{2} g^2 \right) J_1 (\xi ,0),
\end{equation}
where we used the notation:
\begin{equation}\label{J_n}
J_n(\xi,ma_s)=\frac{a_s^{4-2n}}{(2\pi )^4 }\prod_{\rho =1}^4 \int_{
-\pi/a_\rho}^{\pi/a_\rho}
dk_\rho  {1 \over (m^2+{\hat k}^2)^n}.
\end{equation}
Inserting
\begin{equation}
a_s^2 m_H^2 = -2 \left(\frac{1-2\lambda }{\kappa}\xi -6-2\xi^2\right)
\end{equation}
for the one-loop corrected bare mass and using  the notation 
$\lambda = \kappa^2 g_0/(6 \xi)=
4\kappa^2 \lambda_c /\xi $ together with  $a_s^2 (m_H^R )^2 =0 $, 
we get solving perturbatively for $\kappa$:
\begin{equation}\label{kappa_c}
\kappa_c={\xi \over 2(3+\xi^2)}+
{1 \over (3+\xi^2)^2}\left[6\xi J_1(\xi,0)-{\xi^2 \over (3+\xi^2)} \right]
\lambda_c +{9\xi J_1(\xi ,0) \over 16(3+\xi^2)^2}g^2.
\end{equation}
This result coincides with the $d^2V_{eff}(\Phi=0)/d\Phi^2=0$ condition of 
eq. (12) of \cite{Cs-F}. 
For the readers'
convenience we plot $J_1(\xi,0)$ of eq. (\ref{J_n}) in figure 1 as a
function of $1/\xi$. For the special case  of symmetric lattice spacings, $\xi=1$,
our quantum corrections to the critical
hopping parameter reproduce the known result of the isotropic SU(2)-Higgs
model (\cite{montvay,hasenfratz}).
An 8-term    Chebishev polynomial approximation with $6\cdot 10^{-6}$  
 accuracy 
to the function reads:
\begin{eqnarray}
J_1(\xi,0)= 0.2276734-0.000175561/\xi-0.1452559/\xi^2
\nonumber \\
-0.03593908/\xi^3+0.3487585/\xi^4-0.4128226/\xi^5
\nonumber \\
+0.2187872/\xi^6-0.04609285/\xi^7.
\end{eqnarray} 

It is instructive to check that the same result is obtained starting from 
the symmetric phase perturbation theory, where some graphs are absent and 
one is lead to
\begin{equation}\label{Higgs-mass_sym}
0 =-a_s^2 m_0^2 -\left(g_0+\frac{9}{4} g^2 \right) J_1 (\xi ,0).
\end{equation}

\begin{figure} \begin{center}
\epsfig{file=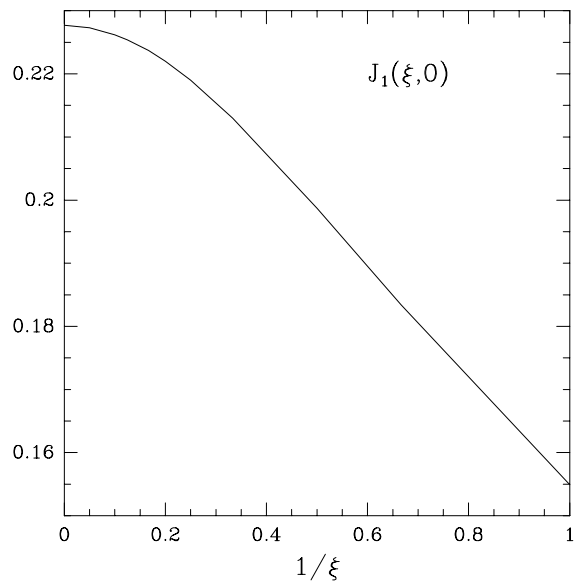,width=8.0cm}
\caption{\label{j_1}
{\sl The lattice integral $J_1(\xi,0)$ (see text) on asymmetric lattices
as a function of $1/\xi$.
}}
\end{center}\end{figure}

Let us now consider the self-energy corrections to the gauge boson mass. 
The relevant diagrams are shown in figure 3.
The inverse propagator (eqs. (\ref{W1-propagator})--(\ref{W3-propagator})) 
at zero momentum has a specific structure, namely 
\begin{eqnarray}\label{W_prop_0}
\tilde \Delta_{W,ij}^{ab} (0)^{-1} =\delta^{ab} \delta_{ij} m_W^2 
\frac{g_s^2}{g^2}, \nonumber \\
\tilde \Delta_{W,44}^{ab} (0)^{-1} =\delta^{ab} m_W^2
\frac{g_t^2}{g^2} \frac{\gamma_\kappa^2}{\xi^2}.
\end{eqnarray}
One therefore has to determine both the diagonal space--space and the 
time--time  components in order to check consistency. Since the bare mass 
squared  turns out to be ${\cal O}(g^2)$, we may safely put 
$g_s^2/g^2=g_t^2/g^2=\gamma_\kappa^2/\xi^2=1$ in 
(\ref{W_prop_0}). Finally we obtain, after imposing zero renormalized 
lattice unit mass squared:
\begin{eqnarray}\label{W_mass}
a_s^2 m_W^2 =g^2 \left(\frac{3}{2} +\frac{9}{2}\frac{m_W^2}{m_H^2} 
\right) J_1 (\xi ,0).
\end{eqnarray}
Inserting 
\begin{eqnarray}
m_W^2= m_H^2 \frac{3 g^2}{4 g_0},
\end{eqnarray}
we get back eq. (\ref{kappa_c}) consistently. 
Again we have checked that (\ref{W_mass})
holds in all $R_\xi $ gauges.

Next we discuss the anisotropy parameters $\gamma_\beta$ and $\gamma_\kappa$.  
Following Karsch and Stamatescu \cite{karsch89} we determine them 
 from the requirement of rotational invariance in the continuum limit 
$a_s , a_t \rightarrow 0$ at fixed $\xi =a_s /a_t $.
In particular we consider the physical particle propagators, which receive 
quantum corrections
\begin{equation}\label{propagators}
\tilde\Delta_{H,1}(p)^{-1}=\tilde\Delta_H(p)^{-1}+\Sigma_{H,1}(p),\ \ \
\tilde\Delta_{W,1,\mu\nu}^{ab}(p)^{-1}=
\tilde\Delta_{W,\mu\nu}^{ab}(p)^{-1}+
\Sigma_{W,1,\mu\nu}^{ab}(p),
\end{equation}
where $\tilde\Delta_H(p)^{-1}$ and $\tilde\Delta_{W,\mu\nu}$ 
 are the tree-level propagators corrected with the two-point vertices. 
 $\tilde\Delta_H(p)^{-1}$ is given by
\begin{equation}\label{H1-propagator}
\tilde\Delta_H(p)^{-1}=m_{H,0}^2+\sum_{i=1}^3 {\hat p}_i^2
+{\gamma_\kappa^2 \over \xi^2}{\hat p}_4^2,
\end{equation}
while $\tilde\Delta_{W,\mu\nu}$ is given by 
(\ref{W1-propagator})--(\ref{W3-propagator}).

\vspace{2cm}

\begin{figure}[htb]
\begin{center}
\epsfig{file=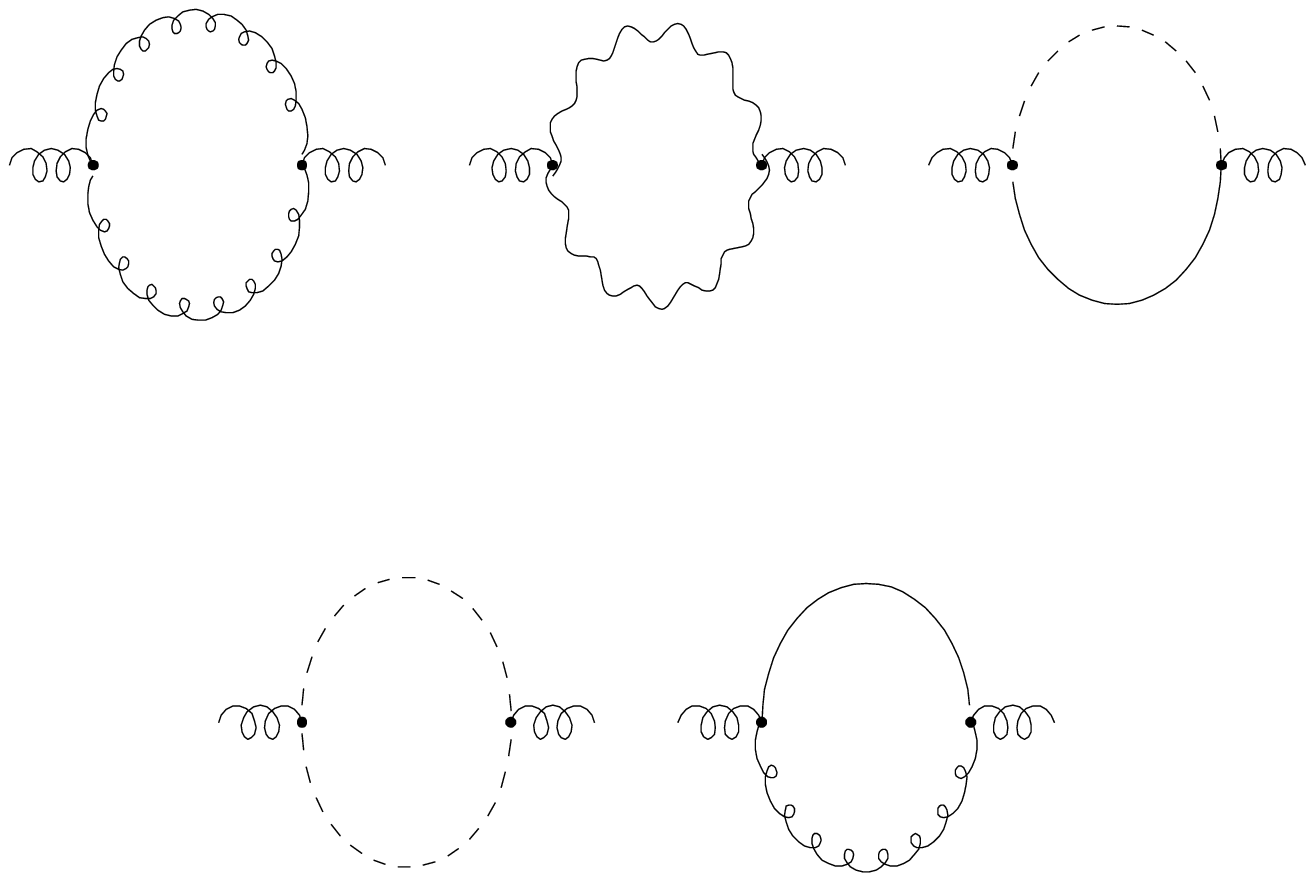,bbllx=150pt,bblly=400pt,bburx=480pt,bbury=680pt,
width=6.4cm}
\epsfig{file=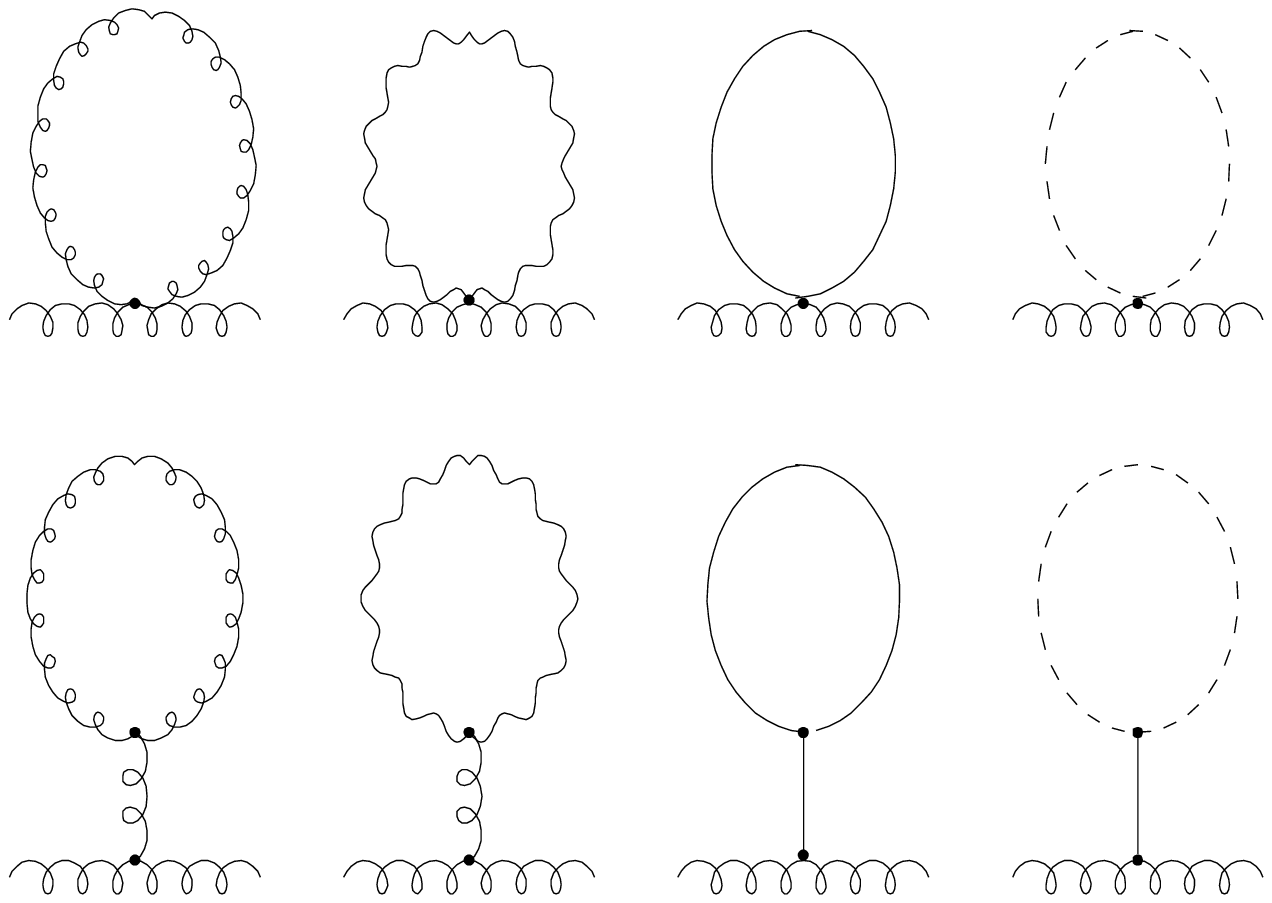,bbllx=60pt,bblly=340pt,bburx=480pt,bbury=620pt,
width=8cm}
\caption{\label{Vector_self}
{\sl Vector boson self energy graphs.
 Solid lines stand for Higgs, dashed
lines for Goldstone, curly lines for vector bosons and wavy lines denote ghosts.
}}
\end{center}\end{figure}

The corrections to the anisotropies in the kinetic parts of
eqs. (\ref{propagators}) should be cancelled
by the kinetic parts of the self-energies.
For the Higgs boson this can be achieved by requiring 
\begin{equation}\label{H_anisotropy}
1+\frac{1}{2}\frac{\partial ^2 \Sigma_{H,1}(p)}
{\partial p_i^2}\Big|_{p=0}=\frac{\gamma_\kappa^2}{\xi^2}
+\frac{1}{2}\frac{\partial ^2 \Sigma_{H,1}(p)}
{\partial p_4^2}\Big|_{p=0},
\end{equation}
where $i=1,2,3$. The graphs contributing are 
the momentum dependent ones of figure 1.

For the gauge boson there are several possibilities. As a simple one we 
choose 
\begin{equation}\label{W_anisotropy}
1+\frac{1}{2}\frac{\partial ^2 \Sigma_{W,1,ii}(p)}
{\partial p_j^2}\Big|_{p=0}=\frac{\gamma_\beta^2}{\xi^2}
+\frac{1}{2}\frac{\partial ^2 \Sigma_{W,1,ii}(p)}
{\partial p_4^2}\Big|_{p=0},
\end{equation}
where $i \neq j=1,2,3$. This is easily calculated, since   only the  
$\delta_{i,j} $ 
term of $\tilde\Delta_{W,ij}^{ab}(p)^{-1}$ contributes on the left hand side. 
Not all the self energy graphs contribute, but only those 
graphs of figure 3, which depend on the momentum. 

Our results for infinite lattices are
\begin{equation}\label{res_b}
b_\beta (\xi) =0, \ \ b_\kappa (\xi) =0,
\end{equation}
\begin{eqnarray}\label{res_c_beta}
c_\beta (\xi)=
\int_{-\frac{\pi}{a_s}}^{\frac{\pi}{a_s}}
\int_{-\frac{\pi}{a_s}}^{\frac{\pi}{a_s}}
\int_{-\frac{\pi}{a_s}}^{\frac{\pi}{a_s}}
\int_{-\frac{\pi}{a_t}}^{\frac{\pi}{a_t}} d^4 q 
 \frac{a_s^2}{\sum \hat q_\mu^2} 
\left \{ \left[\bigg(1-\frac{1}{\xi^2}\bigg )\frac{3+\cos (q_1 a_1)}{4}-
\frac{\cos (q_3 a_3)-\cos (q_4 a_4 )}{4}\right]
 \right. \nonumber \\ \left.
+\frac{1}{a_s^2 \sum \hat q_\mu^2}\left(-\bigg 
(1-\frac{1}{\xi^2}\bigg )\sin ^2 (q_1 a_1 )
+\frac{9}{2} \cos ^2 (\frac{q_1 a_1 }{2}) (\cos (q_3 a_3 )-\cos (q_4 a_4 ))
\right)
 \right. \nonumber \\ \left.
+\left(\frac{8 \sin ^2 (q_1 a_1 )}{(a_s^2 \sum \hat q_\mu^2)^2}+\frac{2 \cos ^2 
(\frac{q_1 a_1 }{2})}{a_s^2 \sum \hat q_\mu^2 } \right)
\left(\frac {\sin ^2 (q_3 a_3 )-\xi^2 \sin^2 (q_4 a_4 )}{a_s^2 \sum \hat q_\mu^2 }
-\frac{\cos (q_3 a_3 )-\cos (q_4 a_4)}{2}\right)
 \right. \nonumber \\ \left.
+\frac{\sin ^2 (q_1 a_1 )}{(a_s^2 \sum \hat q_\mu^2 )^2} 
\left(\frac{ \sin ^2 (q_3 a_3 )-
\xi^2 \sin ^2 (q_4 a_4 )}{a_s^2 \sum \hat q_\mu^2 }-
\cos (q_3 a_3 )+\cos (q_4 a_4 )\right)\right \},
\end{eqnarray}

\begin{eqnarray}\label{res_c_kappa}
c_\kappa (\xi)=\frac{3}{4} \int_{-\frac{\pi}{a_s}}^{\frac{\pi}{a_s}}
\int_{-\frac{\pi}{a_s}}^{\frac{\pi}{a_s}}
\int_{-\frac{\pi}{a_s}}^{\frac{\pi}{a_s}}
\int_{-\frac{\pi}{a_t}}^{\frac{\pi}{a_t}} d^4 q \frac{a_s^2}{\sum \hat q_\mu^2}
\left \{\bigg (1-\frac{1}{\xi^2}\bigg 
 )\frac{16}{(a_s^2 \sum \hat q_\mu^2 )^2 }
\left(\xi^2 \sin^2 (q_4 a_4 )-\sin^2 (q_1 a_1 )\right)\right\},
\end{eqnarray}
where the sums are over $\mu =1, \dots,4 $.
The above expressions are easily seen to be finite and independent of 
$a_s$ and $a_t$. We have also checked that they are gauge independent. 
The dependence on $\xi$ is plotted in figure 4.

A 6-term    Chebishev polynomial approximation with $2\cdot 10^{-5}$ 
accuracy to the functions reads:
\begin{eqnarray}
c_\beta (\xi)= -0.1687249+0.124013/\xi+0.08608489/\xi^2 
\nonumber \\
-0.04715295/\xi^3 -0.0002526438/\xi^4+0.006038775/\xi^5  , 
\end{eqnarray}
\begin{eqnarray}
c_\kappa (\xi)=  -0.05691582-0.0001275536/\xi+0.07582766/\xi^2  
\nonumber \\
-0.003112956/\xi^3 -0.0265274/\xi^4+0.01085953/\xi^5.
\end{eqnarray}

We also have to equate  $\tilde\Delta_{W,1,13}$ and 
$\tilde\Delta_{W,1,14}$:
\begin{equation}\label{W13=W14}
1+\frac{g_s^2}{\alpha g^2}-\frac{\partial^2 \Sigma_{W,13}}{\partial p_1 
\partial p_3}
\Big|_{p=0}=
\frac{\gamma _\beta}{\xi}+\frac{g_s g_t}{\alpha g^2}\frac{\gamma_\kappa^2}{\xi}
-\frac{\partial^2 \Sigma_{W,14}}{\partial p_1 \partial p_4}\Big|_{p=0}.
\end{equation}
This is a non-trivial constraint, which our previous expressions do satisfy.

\begin{figure} \begin{center}
\epsfig{file=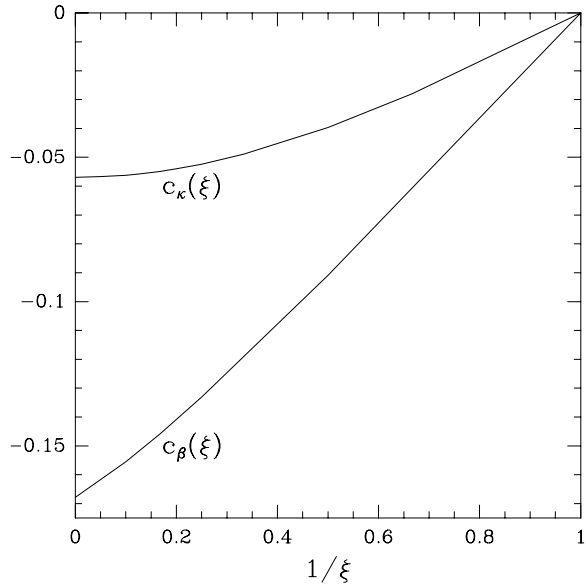,width=8.0cm}
\caption{\label{vector}
{\sl $c_\beta (\xi)$ and $c_\kappa(\xi)$ as functions of
$1/\xi$.}}
\end{center}\end{figure}

There are several important features of the anysotropy parameter result, which
should be mentioned.\hfill\break
{\it a. Masses in the propagators:} A consistent perturbative procedure
on the lattice determines the bare parameters, for which the renormalized
masses vanish, cf. eq. (\ref{kappa_c}). With these bare couplings other
quantities, e.g. asymmetry parameters, are determined. However, using
the one-loop renormalized masses ($a_s m_H^R=a_s m_W^R=0$)
in the propagators instead of the bare ones leads to changes in the results,
which are higher order in  $g^2$ and $\lambda$. Therefore, all our
results are given by the integrals with renormalized masses.
\hfill\break
{\it b. $g^2$ and $\lambda$ corrections:} In figure 4 we have
given only $c_\beta (\xi)$ and $c_\kappa(\xi)$. As shown by 
eq. (\ref{res_b}) the functions 
$b_\beta (\xi)$ and $b_\kappa(\xi)$ vanish, thus there are no
corrections of ${\cal O}(\lambda)$ to the anisotropy parameters.
It is easy to understand this result qualitatively, since only
graphs with two or more scalar self-interaction vertices have non-trivial
dependence on the external momentum. This feature is connected with
the well-known fact that the $\Phi^4$ theory does not have any
wave function correction in first
order in the scalar self-coupling. It is worth
mentioning that there is
only one type of two-loop graph (the setting-sun) which should be
combined with the one-loop graphs, in order to obtain the whole
${\cal O}(\lambda^2)$ correction.\hfill\break
{\it c. Pure gauge theory:} 
A number of graphs of figure 3  (namely those containing only vector boson 
and ghost lines) are identical to those of the pure 
gauge theory. Evaluating the momentum dependent ones from 
these diagrams, one reproduces the result of ref. \cite{karsch82}
(the function $c_\beta (\xi)$ of the present paper corresponds to
$c_\tau(\xi)-c_\sigma(\xi)$ of ref. \cite{karsch82}).
The most important contribution comes from the self-energy graph with
gauge boson four-coupling.
Inclusion of the scalar particles
gives only small changes. The relative difference between the $c_\beta (\xi)$
functions for the pure SU(2) theory and for the SU(2)-Higgs model
is typically a few \%.\hfill\break
{\it d. Quantum corrections to the hopping parameter:} The
contributions to the hopping parameter come from 
the momentum dependent graphs of figure 1. 
This correction has the same sign and order of magnitude than that of
the gauge anisotropy parameter; however it is somewhat smaller. It is
possible to combine the anisotropies
$c_\beta '(\xi)=c_\beta (\xi)-c_\kappa(\xi)$.
For this choice in the gauge sector and with $\gamma_\kappa =\xi$ the
rotational invariance can be restored on the one-loop level, choosing the
appropriate value for the lattice spacing asymmetry $a_s /a_t$. Thus, the
masses in both directions will be the same. However, the obtained lattice
spacing asymmetry will then slightly differ from the original $\xi$.
One gets $a_s/a_t= \xi(1-g^2c_\kappa(\xi)/2)+{\cal O}(g^4,\lambda^2)$.\\
{\it e. For later use we specify:} $c_\beta (4)=-0.13308$, 
$c_\kappa (4)=-0.052353$,
thus $c_\beta '(4)=-0.080727$, $\gamma_\beta ' (4)=3.9193$, $a_s /a_t =
4.05235$.\\
{\it f. Asymmetry parameters away from the critical line:}
Following the procedure outlined above one may determine the asymmetry 
parameters away from the critical line. In this case tree-level masses are 
nonvanishing and are in fact ${\cal O} (1)$. Therefore one has to keep them 
in the 
propagator denominators. Thus the final results become more complicated. 
We do not reproduce the formulae here, only note that 
 numerically the results are very close to the previous case. Thus 
the asymmetries determined near the critical line are universally applicable.\\
{\it g. Finite lattice results:}
The above formulae are valid for infinite lattice sizes, however, replacing 
the lattice integrals with the appropriate lattice sums, one gets  
results valid for finite lattices.

\subsection{Perturbative study of the continuum limit of the finite temperature 
 theory 
and optimal choice of the parameter $\xi$}

The approach to the continuum  limit of the finite temperature theory 
may be studied in the approximation 
of one-loop perturbation theory. The relevant physical quantities we study 
are the ratio of the critical temperature ($T_c$) and the Higgs mass and 
the ratio of the Higgs and vector boson masses. 
To calculate them in perturbation theory we first determine the  
bare Higgs mass parameter using  the analogue of eq. (\ref{Higgs-mass}) for a 
 lattice with finite extension ($L_t$) in the $t$-direction, 
 i.e. at finite temperature $T=1/(L_t a_t )$, by imposing the  
condition $a_s^2 (m_H^R)^2 = 0$. This choice corresponds to the lowest point  
of the metastability    region with $T_c=1/(L_t a_t )$, i.e. when the 
derivative of the effective potential at zero field first becomes negative. 
Using the same bare coupling parameters in the action we 
next determine the physical Higgs and vector 
boson masses on a T=0  lattice (i.e. using a lattice with equal (infinite) 
physical dimensions in space and time directions).

More precisely, the bare   quantity $a_s^2 m_H^2$ is determined from Eq. 
(\ref{Higgs-mass}) with $a_s^2 (m_H^R)^2=0$, replacing however 
$J_1 (\xi,0)$ with $J_T (L_t ,\xi,0)$, where 

\begin{equation}\label{J_T}
J_T(L_t ,\xi,ma_s)=\frac{\xi}{L_t}\sum_{n_t=0}^{L_t-1}\frac{a_s}{(2\pi )^3 }\prod_{\rho =1}^3 \int_{
-\pi/a_s}^{\pi/a_\rho}
dk_\rho  {1 \over (m^2+{\hat k}^2)},
\end{equation}
and   in the denominator $\hat{k}_4 $ is given by
\begin{equation}\label{k_4}
\hat{k}_4 =\frac{2}{a_t}\sin \frac{2\pi n_t}{L_t}.
\end{equation}
(It is straightforward to write down the finite lattice version of eq. 
(\ref{J_T}), too.)
The $T=0$ renormalized Higgs mass ($a_s m_H^R $) is then determined 
from the unmodified eq. (\ref{Higgs-mass}) using the 
already known value of the bare parameter $a_s^2 m_H^2$ and the infinite 
volume $T=0$ integral $J_1 (\xi,0)$. 
Using $T_c =1/(a_t L_t ) =\xi/ (a_s L_t )$ we finally obtain the simple 
formula for a given $L_t $:
\begin{equation}\label{pertT_cm_H}
T_c/m_H^R=\frac{\xi}{L_t}\frac{1}{[(2g_0+9g^2/2)\,
(J_T(L_t ,\xi,0)-J_1(\xi,0))]^{\frac{1}{2}}}.
\end{equation}
In the same approximation $(m_H^R /m_W^R)^2 $ equals to the tree level value
 $4g_0/(3g^2)$.

\begin{figure} \begin{center}
\epsfig{file=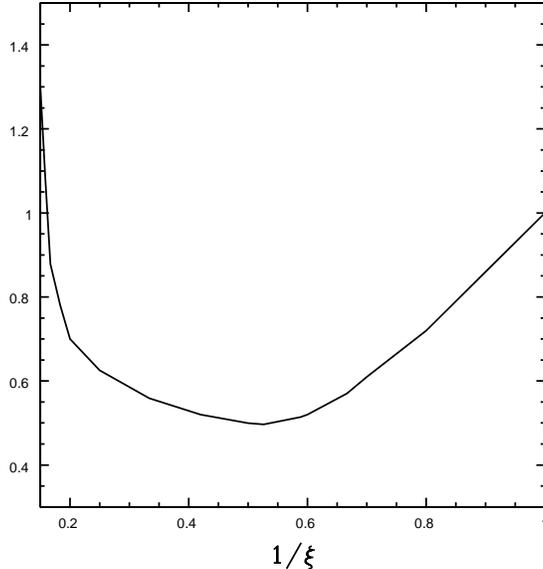,height=8cm}
\caption{\label{opt}
{\sl 
Simulation time  
necessary to reach 0.1\% precision determination of 
$T_c$/$m_H^R$ (normalized to the   $\xi=1$ point) versus $1/\xi$.
}}
\end{center}\end{figure}

The result eq. (\ref{pertT_cm_H}) refers to infinitely large lattices (i.e. 
infinite in both the space-like and time-like directions for the $T=0$ case and 
infinite in only the space-like direction for the $T \neq 0$ case.)
The continuum limit (realized as   $L_t \rightarrow \infty$) is well defined.
In lattice simulations, however, we always have finite lattices. We have to 
choose minimal lattice volumes large enough to ensure a reasonable precision.
 This choice 
of course does depend on $\xi$, therefore we may also look for the 
optimal choice  of $\xi$ ensuring a reasonable precision (say 0.1\%) of the 
physical mass determinations using the smallest possible lattices or 
shortest simulation times.
This problem may be studied in lattice perturbation theory.

To obtain the optimal choice of $\xi$ we first determine the 
$L_t \rightarrow \infty$ (i.e. the continuum) limit value of  $T_c/m_H^R$ as 
a function of $\xi$ using eq.
 (\ref{pertT_cm_H}). 
We obtain that -- as expected -- the limit of $T_c/m_H^R$ does not depend 
on $\xi$  within errors. 

Next we take into account that in practice we simulate on lattices with 
finite extensions. 
In order to fit in  the relevant modes we have to deal with a given
physical volume:
\begin{equation}\label{sim_vol}
V_{phys}=L_t a_t (L_s \xi a_t)^3= {1 \over T} (L_s/L_t \cdot \xi/T)^3 .
\end{equation}
Thus the number of the lattice points (which determines the memory required) 
 is expressed as 
\begin{equation}\label{memory}
L_t L_s^3=V_{phys} T^4 L_t^4/\xi^3 .
\end{equation}
To get a correct estimate of the simulation time we have to take into account 
the autocorrelation times as well. 
 Since these are
proportional to the squares of the correlation lengths for a local 
updating algorithm (see \cite{Wolff}), i.e. to $(L_t/\xi)^2$, the time 
necessary for simulation on a given physical volume and temperature  
will be proportional to 
\begin{equation}\label{sim_time}
V_{phys} T^4 L_t^6/\xi^5 .
\end{equation}
Next we choose a lattice extension in temporal direction $\bar{L_t}$ so that by eq.
(\ref{pertT_cm_H}) we obtain an approximation of  the previously determined 
continuum limit  
$T_c/m_H^R$ value to a given (say 0.1\%) precision. 
 $T_c/m_H^R$ as determined from eq. (\ref{pertT_cm_H}) as a
function of $L_t$ approaches the limiting value from below for large $L_t$
 for all $\xi$ values. However,
for $\xi \ge 2$ it decreases for increasing, small $L_t$ values.
Thus specific small $L_t$ values may better approximate the limiting value 
of $T_c/m_H^R$ than larger 
intermediate values. It is clear that this is an accidental agreement only,
therefore in our considerations we have determined the smallest $\bar{L_t}$ 
value giving $T_c/m_H^R$ with  the required precision, which does not
deteriorate for larger $L_t$. 

More precisely we compare the true 
continuum limit  of $T_c/m_H^R$ with an approximate value obtained 
 from an extrapolation to $L_t=\infty$ 
of the $T_c/m_H^R$ values determined from four subsequent $L_t$ values. 
We choose $\bar{L_t}$ to be the minimal $L_t$, which (together with 
the 3 larger $L_t$ values) already gives the required precision. 
Having determined $\bar{L_t}$ we calculate the corresponding simulation time 
for finite lattice size using eq. (\ref{sim_time}).
 Figure 5 shows the simulation time 
normalized to the $\xi=1$ value as a function of $1/\xi$ for 0.1\% 
precision in $T_c/m_H^R$. 
The normalized simulation time as a function of $\xi$ has a broad minimum 
near $\xi=2$. The number of lattice points (\ref{memory})
 (normalized to the $\xi=1$ value) is quite a similar function of $\xi$ 
with a broad minimum near $\xi=2$.
In our numerical simulations we have choosen  $\xi \simeq  4 $, which is a 
good choice both from the point of view of simulation time and fiting in the 
relevant modes into a practically accessible lattice.

\section{Non-perturbative analysis of the anisotropies}

This  section of the paper deals with our non-perturbative determination
of the anisotropy parameters by means of numerical simulations.
Besides a mere confirmation of the one-loop calculations in the previous
part, it could give estimates of possible corrections, which go beyond 
perturbation theory.
This is an important step towards future studies of the finite temperature 
electroweak phase transition in the framework of the four-dimensional 
SU(2)-Higgs model on anisotropic lattices.
Namely, if the deviation from the perturbative results turns out to be so 
small that its influence on expectation values in a numerical simulation is 
negligible within their typical statistical errors, the one-loop 
perturbative anisotropies $\gb$, $\gk$ and $\xi$ can be used without any 
further (non-perturbative) fine-tuning.
At first sight this may not seem very surprising, because the 
zero-temperature theory is weakly coupled ($g^2\simeq 0.5$).
But owing to the fact that the corrections in the parameter $\lambda$ 
--- entering only at two-loop level --- whose size essentially determines
the value of the Higgs boson mass, are not exactly known, such an 
investigation is necessary, particularly in view of Higgs 
masses around 80 GeV or larger, which is the physically allowed 
region determined by the LEP experiments.

As already discussed above, the tree-level values of the anisotropies 
receive quantum corrections, which in general have to be determined 
non-perturbatively.
A physically motivated idea for their estimation is to impose the 
restoration of the space-time interchange symmetry as a remnant of Lorentz
invariance after discretization of the continuum theory.  
In practice this is to be realized by the requirement that Higgs and gauge  
boson correlation lengths in physical units should be equal in space- and 
time-like directions.
Furthermore, we include into the analysis the length scale of the static 
potential derived from space-time and space-space Wilson loops.
 
The following subsections describe our numerical studies in more detail.
After some brief remarks on the simulation techniques and parameters used, 
we present the results on the physical observables under consideration and 
propose, how they can serve to extract the coupling and lattice spacing
anisotropies non-perturbatively.
Finally, the values obtained in this way are confronted with perturbation 
theory.   
\subsection{Monte Carlo simulation and its parameters}
In our Monte Carlo simulations we apply an optimized combination of heatbath
and overrelaxation algorithms, which has been extensively discussed for the 
isotropic model in refs. \cite{FHJJM95,CFHJM96,HH96}, and their 
implication carries over straightforwardly to an anisotropic lattice.
The action (\ref{lattice_action}) is easily arranged to 
$S[U,\varphi]=\sum_{x\in\Lambda}S_x$, 
and the lattice action per point
\be
S_x=6\beta\plaqx+R_x+\lambda Q_x-8\kappa\lphix
\label{actionpp}
\ee
consists of the length variables of the Higgs field 
\be
R_x\equiv\half\tr{\hf{\varphi}^+\hf{\varphi}}=\hf{\rho}^2\,,\quad
Q_x\equiv(\hf{\rho}^2-1)^2\,,
\label{lengvars}
\ee
of the weighted sum of the plaquette contributions 
$U_{x;\mu\nu}\equiv
\gf{U}{\mu}\gfup{U}{\mu}{\nu}\gfup{U}{\nu}{\mu}^+\gf{U}{\nu}^+$
lying in the space-space and the space-time planes
\be
\plaqx=
\frac{1}{6}\left(\frac{3}{\gb}\,\plaqsx+3\gb\plaqtx\right)
\label{plaqwght1}
\ee
\be
\plaqsx\equiv
\frac{1}{3}\sum_{1\le\mu<\nu\le3}\Big(1-\half\Tr U_{x;\mu\nu}\Big)
\,,\quad
\plaqtx\equiv
\frac{1}{3}\sum_{\mu=1,3\,;\,\nu=4}\Big(1-\half\Tr U_{x;\mu\nu}\Big)\,,
\label{plaqwght2}
\ee
and of the weighted sum of the space- and time-like components of the 
$\varphi$--link operator 
$\lkop{L}{\varphi}{x\mu}\equiv
\half\tr{\hfup{\varphi}{\mu}^+\gf{U}{\mu}\hf{\varphi}}$:
\be
\lphix=
\frac{1}{4}\left(\frac{3}{\gk}\,\lphisx+\gk\lphitx\right)
\label{lphiwght1}
\ee
\be
\lphisx\equiv
\frac{1}{3}\sum_{\mu=1}^3\lkop{L}{\varphi}{x\mu}\,,\quad
\lphitx\equiv\lkop{L}{\varphi}{x4}\,.
\label{lphiwght2}
\ee
For $\xi=\gb=\gk=1$ this action simplifies to its well known form on 
isotropic lattices.
Eqs.~(\ref{actionpp}) -- (\ref{lphiwght2}) already cover most of the 
observables, whose expectation values are calculated by 
numerical simulations.
 
The updating scheme per sweep, a sequence of one $\gf{U}{\mu}$-- and 
one $\hf{\varphi}$--heatbath step, succeeded by one $\gf{U}{\mu}$-- and 
three $\hf{\varphi}$--overrelaxation steps, has been taken over from 
refs.~\cite{CFHJM96,HH96}.
There it was observed that the inclusion of the 
overrelaxation  algorithms \cite{B95}
reduced the autocorrelation times substantially, in particular for the
operators $\rho^2$ and $\lphi$, whose expectation values show the largest
autocorrelations.

As pointed out in the introduction, the anisotropic version of the 
SU(2)-Higgs model is believed to provide quantitative insights into the
electroweak phase transition at large Higgs boson masses of 
$\mh\ge 80$ GeV, at which the typical excitations with small masses 
(i.e.~large correlation lengths) would demand very large isotropic lattices 
exceeding any presently accessible computer resources. 
In principle a rough resolution in the spatial directions by moderate 
lattices combined with accordingly large lattice anisotropies $\xi$ could 
handle this situation.
However, for $T>0$ a small temporal extension $L_t$ sets the (very large) 
temperature scale through $T=1/a_tL_t$, and hence it is more sensible to 
ensure a large enough lattice cutoff by employing $\xi\simeq L_t$, 
thus~in our numerical work we take
\be
\xi \simeq  4\,, 
\label{xisimval}
\ee
 which is also strongly motivated  by the result  of subsection III.C.
Since this makes the correlation lengths in time direction smaller than
in space directions, it seems to be reasonable to fulfill 
$L_t\simeq \xi L_z$ in order to restore the symmetry of the physical 
extensions and to enable a precise mass determination.
We consider two lattices of sizes $8^2\times12\times48$ and
$8^2\times16\times64$, where the spatial correlation lengths 
correspond to few lattice units  and the finite-volume effects are 
expected to be small.

The $T=0$ simulations are generically intended to fix the physical 
parameters, i.e.~renormalized couplings and masses.
Consequently, the lattice parameters in this study are chosen to reach the 
interesting region of $\mh\simeq 80$ GeV or a Higgs to gauge  boson mass 
ratio of
\be
\hw\equiv\frac{\mh}{\mw}\simeq 1
\label{mratiodef}
\ee
with the experimental input $\mw=80$ GeV setting the overall physical scale.
This is (at least approximately) achieved by the values $\beta=8.0$ and 
$\lambda=0.000178$.
The scalar hopping parameter, which has to comply with the condition that 
the $T>0$ system is at a phase transition point for a certain temporal 
lattice extension, is calculated from  the discretized version of 
eq.~(\ref{kappa_c})\footnote{
The knowledge of the more accurate, non-perturbative value of the critical
hopping parameter, which has to be determined numerically, is not relevant
here.
}.
Referring to $L_t=4$ this amounts to $\kappa=0.10662$.
The non-perturbative corrections usually tend to decrease the tree-level 
mass ratio
\be
\hwtr\equiv\frac{\mhtr}{\mwtr}=
\sqrt{\frac{2\lambda\xi\beta}{\kappa^2}}\,.
\label{mratiotree}
\ee

Our strategy for the determination of the coupling anisotropies is as follows.
In the numerical simulation we have to find those couplings of eq. 
(\ref{lattice_action}), for which the space--time symmetry is restored. 
Therefore, we fix one of the coupling anisotropies to its tree-level value,
ignoring its quantum corrections, and tune the other one to produce 
identical ratios of (decay) masses in space- and time-like directions for 
a set of two or more (particle) channels.
The mass ratios determine the actual lattice anisotropy, which 
will then slightly differ from the 
original $\xi$ of (\ref{xisimval}). 
In this spirit we choose three pairs of coupling anisotropies, denoted as
`tree', `low' and `perturbative',
\bea
\mbox{t :}& \quad &\gk=4.0\,,\quad\gb=4.0\nonumber\\
\mbox{l :}& \quad &\gk=4.0\,,\quad\gb=3.8\nonumber\\
\mbox{p :}& \quad &\gk=4.0\,,\quad\gb=3.919\,,
\label{cplchoice}
\eea
and calculate the corresponding lattice spacing anisotropies from different
physical quantities as described comprehensively in the subsequent subsections.
Assuming that they depend linearly on $\gb$ in this small interval, we can 
interpolate to a matching point $\big(\uts{\gb}{(np)},\uts{\xi}{(np)}\big)$, 
at which all $\xi$--values coincide within errors.  
These estimates are quoted as our non-perturbative results.

All numerical simulations have been done independently on the APE-Quadrics 
computers at DESY-IfH in Zeuthen, Germany, and --- to a smaller extent ---
on the CRAY Y-MP8 and T90 of HLRZ in J\"ulich, Germany, which offer 64--bit 
floating point precision.
In contrast to some quantities, e.g.~the critical hopping parameter in
$T>0$ simulations, the 32--bit arithmetics of the APE-Quadrics is sufficient 
for the calculation of all $T=0$ quantities, especially for 
particle masses and the static potential.
\subsection{Correlation functions and masses}
We now turn to the determination of the Higgs and gauge  boson masses.
As in refs.~\cite{FHJJM95,CFHJM96}, they were obtained from suitable
correlation functions of gauge invariant, local operators 
integrated over time (space) slices. 
Those are $R_x$ and $\lkop{L}{\varphi}{x\mu}$ for the Higgs mass, and the 
composite link fields
\be
W_{x;rk}\equiv
\frac{1}{2}\,\tr{\tau_r\hfup{\alpha}{k}^+\gf{U}{k}\hf{\alpha}}\,,\quad
\mbox{$\tau_r$: Pauli matrices}\,,\quad
r,k=1,2,3
\label{mwopdef}
\ee
for the gauge  ($W$--boson) mass.

The connected correlation functions $\Gamma_O$ of these operators have 
been measured in the time-like and in one space-like direction.
For the Higgs mass $\mh$ the functions $\Gamma_O(t)$ and $\Gamma_O(z)$
were calculated from $t$-- and $z$--slice averages of $R_x$ and the weighted 
$\varphi$--link $\lphix$ of eq.~(\ref{lphiwght1}).
Since these functions can not be regarded as uncorrelated, we have averaged 
them --- after an appropriate normalization of the correlations at distance 
zero --- before performing the mass fits.    
The same prescription holds for the gauge boson mass $m_W$, but with two major 
differences:
firstly, the $t$-- and $z$--slice correlation functions of $W_{x;rk}$ 
have been measured separately for all combinations of $r$ and $k$, and
secondly, in place of $k=3$ in (\ref{mwopdef}) actually we have to take 
$k=4$ for the correlations in $z$--direction (i.e. all directions in 
$W_{x;rk}$ are orthogonal to the direction of propagation).
Again the individual correlation functions are averaged to one function 
per direction as in the Higgs channel.

As lowest energies  the particle masses are extracted from one-exponential
least squares fits to shapes of the form
\be
\Gamma_O(\ell)=
A\,\Big[\,\Exp^{\,-m\ell}+\Exp^{\,-m(L-\ell)}\,\Big]+C\,,
\quad \ell=0,1,\ldots,\frac{L}{2}\,,
\quad L\in\{L_t,L_z\}
\label{corrshape}
\ee    
with $m\in\{a_t\mht,a_t\mwt\}$ or $m\in\{a_s\mhs,a_s\mws\}$, respectively. 
The constant terms in the vector channel are highly suppressed so that a
two-parameter fit is mostly sufficient.
Each full data sample has been divided into subsamples, and the statistical
errors on the masses originate from jackknife analyses.
All simulation parameters and lattice sizes are collected in 
table~\ref{ParTab}. 

Our fitting procedure consists of correlated fits, sometimes with
eigenvalue smoothing, and simple uncorrelated fits.
For the former we use the Michael-McKerrel method \cite{M94MMc95}, whose
features and application in the SU(2)-Higgs model have been sketched in
ref.~\cite{CFHJM96}.
Its main purpose is to select the most reasonable fit interval in data sets,  
which are strongly correlated in the fitted direction.
Uncorrelated fits, which ignore these correlations,  are often plagued with 
very small values of $\chi^2$ per
degree of freedom (dof) for nearly all fit intervals in question, whereas in
correlated fits the emergence of $\chi^2/\mbox{dof}\simeq 1$ for some fit
intervals represents a safe criterion to select reasonable fit intervals.
This also works well for data sets of lower statistics, if the smallest
eigenvalues of the correlation matrix are smeared via replacing them by 
their average. 
All resulting mass estimates in lattice units are shown in 
tables~\ref{MassTab1} and \ref{MassTab2}.
We chose the largest fit interval with a reasonable $\chi^2/\mbox{dof}$ 
from the correlated fit and the results of the uncorrelated fit along this 
interval as the final fit parameters.
Both fits were always consistent within errors, and other fit intervals 
with comparable or even lower $\chi^2/\mbox{dof}$ did not cause any 
significant changes.
%
\begin{table}[htb]
\begin{center}
\begin{tabular}{|c||c|c|c||c|c|c|c|}
\hline
& & & & \multicolumn{2}{c|}{\rule[-3mm]{0mm}{8mm} correlation functions }
& \multicolumn{2}{c|}{\rule[-3mm]{0mm}{8mm} Wilson loops } \\ \cline{5-8}
  \raisebox{1.5ex}[-1.5ex]{index} & \raisebox{1.5ex}[-1.5ex]{lattice}
& \raisebox{1.5ex}[-1.5ex]{$\gk$} & \raisebox{1.5ex}[-1.5ex]{$\gb$}
& sweeps & subsamp. & subsamp. & indep. sweeps \\
\hline\hline
  t1 & $8^2\times12\times48$ & 4.0 & 4.0   & 100000 & 50  & 50  & 100 \\
  l1 & $8^2\times12\times48$ & 4.0 & 3.8   & 100000 & 50  & 50  & 100 \\
  p1 & $8^2\times12\times48$ & 4.0 & 3.919 & 576000 & 192 & --- & --- \\
\hline
  t2 & $8^2\times16\times64$ & 4.0 & 4.0   & 192000 & 64  & 64  & 150 \\
  l2 & $8^2\times16\times64$ & 4.0 & 3.8   & 192000 & 64  & 64  & 150 \\
  p2 & $8^2\times16\times64$ & 4.0 & 3.919 & 704000 & 256 & 128 & 150 \\
\hline
\end{tabular}
\caption{\label{ParTab} \sl Summary of the numerical simulations
                            for the mass and static potential computations. 
                            The other parameters are $\beta=8.0$, 
                            $\lambda=0.000178$, and $\kappa=0.10662$.} 
\end{center}
\end{table}
%

As emphasized above, the space-time symmetry restoration, which implicitly
establishes $\uts{\xi}{(np)}$, becomes apparent in equal physical 
correlation lengths $a_s =a_t\xi$ of the theory.
Thus we introduce anisotropy parameters in the Higgs and vector channels
by   
calculating the ratios
\be
\xh\equiv\frac{a_s\mhs}{a_t\mht}\,,\quad
\xw\equiv\frac{a_s\mws}{a_t\mwt}
\label{ximasses}
\ee
within the jackknife samples of the space- and time-like masses.
These are displayed again in tables~\ref{MassTab1} and \ref{MassTab2}.
Due to the compatibility of the results from the two lattices one concludes 
that the finite-size effects are quite small.
\begin{table}[htb]
\begin{center}
\begin{tabular}{|c||rl|rl|rl|}
\hline
  \multicolumn{1}{|c||}{\rule[-3mm]{0mm}{8mm} quantity }
& \multicolumn{2}{|c}{\rule[-3mm]{0mm}{8mm} t1 }
& \multicolumn{2}{|c}{\rule[-3mm]{0mm}{8mm} l1 }
& \multicolumn{2}{|c|}{\rule[-3mm]{0mm}{8mm} p1 } \\
\hline\hline
  $a_t\mht$ & $4-18\,:$ & 0.1408(22) & $4-22\,:$ & 0.1370(27) 
& $4-24\,:$ & 0.1387(15) \\ 
  $a_s\mhs$ & $1-6\,:$  & 0.5635(31) & $1-6\,:$  & 0.5611(62) 
& $1-6\,:$  & 0.5603(30) \\
\hline
  \multicolumn{1}{|c||}{\rule[-3mm]{0mm}{8mm} $\xh$ }
& \multicolumn{2}{|c}{\rule[-3mm]{0mm}{8mm} 4.002(67) }             
& \multicolumn{2}{|c}{\rule[-3mm]{0mm}{8mm} 4.097(86) }
& \multicolumn{2}{|c|}{\rule[-3mm]{0mm}{8mm} 4.041(55) } \\
\hline\hline
  $a_t\mwt$ & $8-24\,:$ & 0.1523(13) & $8-22\,:$ & 0.1538(13) 
& $8-24\,:$ & 0.1554(25) \\
  $a_s\mws$ & $1-6\,:$  & 0.6225(29) & $2-6\,:$  & 0.6066(40)  
& $2-6\,:$  & 0.6307(22) \\
\hline
  \multicolumn{1}{|c||}{\rule[-3mm]{0mm}{8mm} $\xw$ }
& \multicolumn{2}{|c}{\rule[-3mm]{0mm}{8mm} 4.091(30) }   
& \multicolumn{2}{|c}{\rule[-3mm]{0mm}{8mm} 3.945(32) }   
& \multicolumn{2}{|c|}{\rule[-3mm]{0mm}{8mm} 4.059(44) } \\ 
\hline\hline
  \multicolumn{1}{|c||}{\rule[-3mm]{0mm}{8mm} $\hwt$ }
& \multicolumn{2}{|c}{\rule[-3mm]{0mm}{8mm} 0.925(15) }
& \multicolumn{2}{|c}{\rule[-3mm]{0mm}{8mm} 0.891(20) }
& \multicolumn{2}{|c|}{\rule[-3mm]{0mm}{8mm} 0.892(18) } \\
  \multicolumn{1}{|c||}{\rule[-3mm]{0mm}{8mm} $\hws$ }
& \multicolumn{2}{|c}{\rule[-3mm]{0mm}{8mm} 0.905(6) }           
& \multicolumn{2}{|c}{\rule[-3mm]{0mm}{8mm} 0.925(12) }           
& \multicolumn{2}{|c|}{\rule[-3mm]{0mm}{8mm} 0.888(7) } \\
\hline
\end{tabular}
\caption{\label{MassTab1} \sl Fit intervals, Higgs and gauge boson masses in 
                              time- and space-like directions, and the 
                              resulting lattice spacing anisotropies 
                              for the smaller lattice.} 
\end{center}
\end{table}
%
%
\begin{table}[htb]
\begin{center}
\begin{tabular}{|c||rl|rl|rl|}
\hline
  \multicolumn{1}{|c||}{\rule[-3mm]{0mm}{8mm} quantity }
& \multicolumn{2}{|c}{\rule[-3mm]{0mm}{8mm} t2 }
& \multicolumn{2}{|c}{\rule[-3mm]{0mm}{8mm} l2 }
& \multicolumn{2}{|c|}{\rule[-3mm]{0mm}{8mm} p2 } \\
\hline\hline
  $a_t\mht$ & $4-32\,:$ & 0.1408(22) & $4-32\,:$ & 0.1370(27) 
& $4-32\,:$ & 0.1378(11) \\ 
  $a_s\mhs$ & $1-8\,:$  & 0.5590(42) & $1-7\,:$  & 0.5586(40) 
& $1-8\,:$  & 0.5550(40) \\
\hline
  \multicolumn{1}{|c||}{\rule[-3mm]{0mm}{8mm} $\xh$ }
& \multicolumn{2}{|c}{\rule[-3mm]{0mm}{8mm} 3.969(73) }             
& \multicolumn{2}{|c}{\rule[-3mm]{0mm}{8mm} 4.078(80) }
& \multicolumn{2}{|c|}{\rule[-3mm]{0mm}{8mm} 4.027(36) } \\
\hline\hline
  $a_t\mwt$ & $8-32\,:$ & 0.1499(31) & $8-30\,:$ & 0.1599(42) 
& $6-32\,:$ & 0.1525(15) \\
  $a_s\mws$ & $1-8\,:$  & 0.6318(40) & $3-8\,:$  & 0.607(11)  
& $2-8\,:$  & 0.6133(27) \\
\hline
  \multicolumn{1}{|c||}{\rule[-3mm]{0mm}{8mm} $\xw$ }
& \multicolumn{2}{|c}{\rule[-3mm]{0mm}{8mm} 4.23(10) }   
& \multicolumn{2}{|c}{\rule[-3mm]{0mm}{8mm} 3.80(13) }   
& \multicolumn{2}{|c|}{\rule[-3mm]{0mm}{8mm} 4.021(48) } \\ 
\hline\hline
  \multicolumn{1}{|c||}{\rule[-3mm]{0mm}{8mm} $\hwt$ }
& \multicolumn{2}{|c}{\rule[-3mm]{0mm}{8mm} 0.940(24) }
& \multicolumn{2}{|c}{\rule[-3mm]{0mm}{8mm} 0.857(26) }
& \multicolumn{2}{|c|}{\rule[-3mm]{0mm}{8mm} 0.904(11) } \\
  \multicolumn{1}{|c||}{\rule[-3mm]{0mm}{8mm} $\hws$ }
& \multicolumn{2}{|c}{\rule[-3mm]{0mm}{8mm} 0.885(8) }           
& \multicolumn{2}{|c}{\rule[-3mm]{0mm}{8mm} 0.921(20) }           
& \multicolumn{2}{|c|}{\rule[-3mm]{0mm}{8mm} 0.905(5) } \\
\hline
\end{tabular}
\caption{\label{MassTab2} \sl The same quantities as in table~\ref{MassTab1} 
for the larger lattice.}
\end{center}
\end{table}
%
%
\subsection{Wilson loops and static potentials}
Another approach to the $\xi$--determination is based on the static 
potential, which has the physical interpretation as the energy of an 
external pair of static charges brought into the system.
To this end we have measured rectangular on-axis Wilson loops 
$\Wij(R_i,R_j)$ of extensions $1\le R_i\le L_i/2$ and 
\mbox{$1\le R_j\le L_j/2$}, lying in space-time and space-space planes.
The gauge configuration was transformed to temporal gauge for space-time 
and to $A_3^r (x)=0$ gauge for space-space Wilson loops, and every loop 
with two sides in $t$-- or $z$--direction, respectively, was included in 
the statistics.

As a generalization of the isotropic lattice case we distinguish between 
static potentials 
\be
\Vij(R_i)=
-\lim_{R_j\rightarrow\infty}\,\frac{1}{a_jR_j}\,\ln\,\Wij(R_i,R_j)
\label{potdef}
\ee
in space-like ($ij=st,ss$) and time-like ($ij=ts$) directions, according to
the $R_j\rightarrow\infty$ extrapolation in the second argument of $\Wij$, 
which is supposed to be done first.
The shape of the potential, which is governed by a massive $W$--boson 
exchange \cite{LMP86}, is known to be Yukawa-like, and calculating 
along the lines of
refs.~\cite{CPT84,HK85} lowest order (tree-level) lattice perturbation
theory yields
\be
\Vij(R_i)=
\frac{3g^2}{2}\,\prod_{n\neq j}\int_{-\pi/a_n}^{\pi/a_n}
\frac{dk_n}{2\pi}\,\frac{\sin^2\left(R_ia_ik_i/2\right)}
{\sum_{n \neq j}\hat{k}_{n}^2+m_{W,0}^2}\,+\,{\cal O}(g^4),
\label{potpert}
\ee
with lattice momenta $\hat{k}_n=2a_n^{-1}\sin\left(a_nk_n/2\right)$, 
$n=1,\ldots,4$.
In the continuum limit this expression reflects the usual screening 
behaviour, i.e.~modulo a constant, 
\be
-\frac{3g^2}{4}\,\frac{\Exp^{\,-m_{W,0} r}}{4\pi r}\,,\quad
r\equiv R_ia_i\,,
\label{potclim}
\ee
independent of $i$ and $j$.
After substituting $p_n=a_nk_n$ with $p_n=2\pi l_n/L_n$ and 
\mbox{$l_n=0,1,\ldots,L_n-1$} on a finite lattice, one obtains from 
eq. (\ref{potpert})
\be
a_i\Vij(R_i)=
\frac{3g^2}{16\pi}\,\Big[\,\iij(\mij,0)-\iij(\mij,R_i)\,\Big]
\,+\,{\cal O}(g^4)\,,
\label{potlatt}
\ee
where $\mij =a_i m_{W,0}$ and 
\be
\iij(\mij,R_i)\equiv
\frac{2\pi}{L_iL_kL_l}\,\sum_{p_i,p_k,p_l}\frac{\cos\left(R_ip_i\right)}
{a_ka_l/a_i^2\,\mij^2+\sum_{n\neq j}4a_ka_l/a_n^2\,
\sin^2\left(\half\,p_n\right)}\,,
\label{lattsum}
\ee
where $k$ and $l$ are different from each other and from $i$ and $j$.

Since $g^2=g_R^2+{\cal O}(g_R^4)$, the simulation results for $V_{ij} $ 
are fitted with the ansatz
\be
a_i\Vij(R_i)=
-\frac{\aij}{R_i}\,\,\Exp^{\,-\mij R_i}+\cij+\dij\gij(\mij,R_i),
\label{yukawafit}
\ee
where $G_{ij} $ is a term correcting for finite-lattice (size and spacing) 
artefacts, and  $A_{ij}$, $\mij$, $C_{ij}$, $D_{ij}$ are the parameters 
to be fitted. $G_{ij}$  reads:
\be
\gij(\mij,R_i)=
\frac{1}{R_i}\,\,\Exp^{\,-\mij R_i}-\iij(\mij,R_i)\,.
\label{lattcorr}
\ee
By definition the "global" renormalized coupling is obtained by identifying 
the coefficient of the contribution relevant at short distances,
\be
g_R^2=\frac{16\pi}{3}\,\aij\,.
\label{rencoupl1}
\ee 
Note that $\mij/a_i$ and also $g_R^2$ as determined from Wilson 
loops with different indices have to be independent of the indices for 
properly choosen coupling anisotropies. 

In a first step of the analysis we performed multi-exponential fits 
$\Wij(R_i,R_j)=\sum_{n=0}^{N}c_n\,\Exp^{\,-V_nR_j}$ in order to get the 
potential for fixed $R_i$ as the ground state energy $V_0$ from the large 
$R_j$ asymptotics of the Wilson loops in (\ref{potdef}).   
Starting at distances $R_j=8-11$ or $R_j=1,2$ in dependence of the 
available range in the fitted direction, a sum of two exponentials gave
always stable fits with an optimal compromise between acceptable 
$\chi^2/\mbox{dof}$ and statistical errors, and with $V_0$ well separated
from higher excitations by a large energy gap.
Subsequently, the resulting potentials\footnote{
More precisely, the potentials have to be rendered dimensionless before,
i.e.~in view of eqs.~(\ref{potdef}) and (\ref{yukawafit}) one has to
attach a factor $a_i/a_j$.
} 
were carefully fitted to eq.~(\ref{yukawafit}), and the values of the best 
fit parameters with its errors from jackknife analyses of the data 
subsamples are listed in table~\ref{PotTab}. 
%
\begin{table}[htb]
\begin{center}
\begin{tabular}{|c||c|c|c|c|c||c|}
\hline
  index & $\aij$ & $\mij$ & $\dij$ & $\cij$ 
& $g_R^2\equiv\frac{16\pi}{3}\,\aij$ & $g_R^2(1/\mij)$ \\
\hline\hline  
  t1, $\Wst$ & 0.0335(12) & 0.626(67)  & 0.044(10)  & 0.0832(4)
& 0.561(19)  & 0.575(38) \\ 
  t1, $\Wts$ & 0.0346(3)  & 0.1479(55) & 0.0401(15) & 0.02763(6)
& 0.5800(43) & 0.605(17) \\
  t1, $\Wss$ & 0.0358(7)  & 0.639(27)  & 0.0238(81) & 0.1105(2)
& 0.600(12)  & 0.592(20) \\
\hline
  l1, $\Wst$ & 0.0354(8)  & 0.593(37)  & 0.0292(68) & 0.0873(4)
& 0.592(14)  & 0.582(28) \\
  l1, $\Wts$ & 0.0351(3)  & 0.1651(41) & 0.0372(7)  & 0.02768(5)
& 0.5881(50) & 0.603(20) \\
  l1, $\Wss$ & 0.0360(7)  & 0.623(29)  & 0.0269(56) & 0.1111(2)
& 0.602(12)  & 0.597(21) \\
\hline\hline
  t2, $\Wst$ & 0.0336(2)  & 0.594(26)  & 0.0332(65) & 0.0833(1)  
& 0.5622(35) & 0.562(15)  \\ 
  t2, $\Wts$ & 0.0343(1)  & 0.1401(19) & 0.0390(9)  & 0.02776(2) 
& 0.5739(14) & 0.5932(78) \\ 
  t2, $\Wss$ & 0.0345(3)  & 0.594(13)  & 0.0346(29) & 0.1110(1)
& 0.5781(54) & 0.5781(72) \\
\hline
  l2, $\Wst$ & 0.0347(2)  & 0.555(19)  & 0.0284(56) & 0.0878(1)
& 0.5821(29) & 0.570(14)  \\
  l2, $\Wts$ & 0.0338(1)  & 0.1429(12) & 0.0342(9)  & 0.02792(2)
& 0.5657(11) & 0.5621(64) \\
  l2, $\Wss$ & 0.0344(4)  & 0.557(16)  & 0.0303(28) & 0.1117(2)
& 0.5761(69) & 0.574(12)  \\
\hline
  p2, $\Wst$ & 0.0339(1)  & 0.576(13)  & 0.0322(36) & 0.0851(1)
& 0.5679(21) & 0.5645(92) \\
  p2, $\Wts$ & 0.0343(1)  & 0.1428(11) & 0.0362(2)  & 0.02780(1)
& 0.5742(12) & 0.5845(50) \\
  p2, $\Wss$ & 0.0345(3)  & 0.5810(98) & 0.0307(19) & 0.1112(1)
& 0.5780(43) & 0.5756(77) \\
\hline
\end{tabular}
\caption{\label{PotTab} \sl All Yukawa fit parameters of the static 
                            potentials, calculated from space-time 
                            ($ij=st$ and $ij=ts$) and space-space ($ij=ss$) 
                            Wilson loops. The renormalized coupling
                            $g_R^2(1/\mij)$ is explained in the text.}
\end{center}
\end{table}
%
We only used uncorrelated fits in the present context, because the size of 
the Wilson loop extensions does not admit much variation in the fit 
intervals.  
In some cases the smallest distances $R_i=1$ or $R_i=1,2$ were omitted
to have a satisfactory $\chi^2/\mbox{dof}$.
This supports the experiences from earlier work \cite{CFHJM96} that the
lattice correction $\gij$ may be not adequate enough for our data.
A more thorough inspection of the fit results hints at a renormalization of 
$g^2=0.5$ on the ${\cal O}(15\%)$--level, and from the validity of 
$\aij\simeq\dij$ one can judge, how good the assumption of a one gauge  
boson exchange really is.
The space-like potentials from $\Wst$ and $\Wss$ lead to consistent numbers,  
but the discrepancy between the screening masses 
$\mij\in\{\msst,\msss,\msts\}$ and the gauge  masses of 
the preceding subsection is often larger than expected.
When comparing the two lattices, we observe only small finite-volume 
effects in $g_R^2$, but the $\mij$ still differ outside their --- even 
larger --- standard deviations.
However, as we will see below, these effects seem to cancel to a great 
extent in the mass ratios we are mainly interested in.

For the sake of completeness we also discuss a local definition of the 
renormalized gauge coupling, which goes back to refs.~\cite{LMP86,S94} and
has been applied to the isotropic SU(2)-Higgs model in
\cite{FHJJM95,CFHJM96}.
Since the short-distance potentials turn out to deviate from a pure Yukawa 
ansatz, we set
\be
g_R^2(R_i)\equiv
\frac{16\pi}{3}\,\frac{a_i\Vij(R_i)-a_i\Vij(R_i-d)}
{\iij(\mij,R_i-d)-\iij(\mij,R_i)}
\label{rencoupl2}
\ee
at distance $R_i$ with $\mij$ as screening masses from the large-distance 
fits to (\ref{yukawafit}).
$R_i$ is the solution of the equation
\be
\frac{1}{R_i}\,\,\Exp^{\,-\mij R_i}\,\left[\,\frac{1}{R_i}+\mij\,\right]=
\frac{\iij(\mij,R_i-d)-\iij(\mij,R_i)}{d}
\label{distdef}
\ee
and is interpolated to the physical scale $\Rij\equiv 1/\mij$, giving the 
typical interaction range of the potential.
Eq.~(\ref{distdef}) is motivated by requiring the force 
$\frac{\dtt}{\dtt R_i}\,a_i\Vij(R_i)$ in the continuum limit (\ref{potclim}) 
to be equal to the finite difference 
$\left[\,a_i\Vij(R_i)-a_i\Vij(R_i-d)\,\right]/d$ as would follow from 
(\ref{potlatt}).
This improves the naive choice $R_i-\,d/2$ to tree-level \cite{S94},
because it compensates for lattice artefacts of order ${\cal O}(a_i^2/r_i^2)$. 
The results for $d=1$ are collected in the last column of table~\ref{PotTab} 
and agree with $g_R^2$ from the global definition. 
The errors contain the statistical errors of the potentials, the
(ever dominating) uncertainties in the masses, and systematic errors by 
accounting for the sensitivity to a quadratic $\Rij$--interpolation with 
three neighbouring points instead of a linear one with only two points.

\newpage
Rotational symmetry now implies that the renormalized gauge coupling and 
$\mij/a_i$ should be independent of $i$ and $j$.
For $g_R^2$ this is obviously true, and in analogy to (\ref{ximasses})
a further kind of lattice spacing anisotropy from the ratios of screening 
masses is
\be
\xp\equiv\frac{\msst}{\msts}\quad\mbox{or}\quad
\xp\equiv\frac{\msss}{\msts}\,.
\label{xiscreen}
\ee
Its values in all simulation points are quoted in table~\ref{XiPotTab}.
In contrast to the masses themselves, they show rather good consistency and
are hardly affected by the finite volume.
%
\begin{table}[htb]
\begin{center}
\begin{tabular}{|c||c|c||c|c|c|}
\hline
  quantity & t1 & l1 & t2 & l2 & p2 \\
\hline\hline
  $\xp=\msst/\msts$  & 4.23(47)  & 3.56(26)  & 4.24(18)
& 3.88(14)                 & 4.033(96) \\
\hline  
  $\xp=\msss/\msts$  & 4.32(24)  & 3.76(20)  & 4.24(12) 
& 3.89(13)                 & 4.068(80) \\
\hline\hline
  $\xp$ via matching       & 4.250(77) & 3.923(62) & 4.179(38) 
& 3.915(52)                & 4.028(31) \\ 
\hline
\end{tabular}
\caption{\label{XiPotTab} \sl Errors for the lattice spacing anisotropy 
                              obtained
                              as ratios of the corresponding screening 
                              masses are computed from their jackknife samples. 
                              The direct matching of the potentials 
                              is described in the text.} 
\end{center}
\end{table}
%

\begin{figure}[htb]
\begin{center}
\epsfig{file=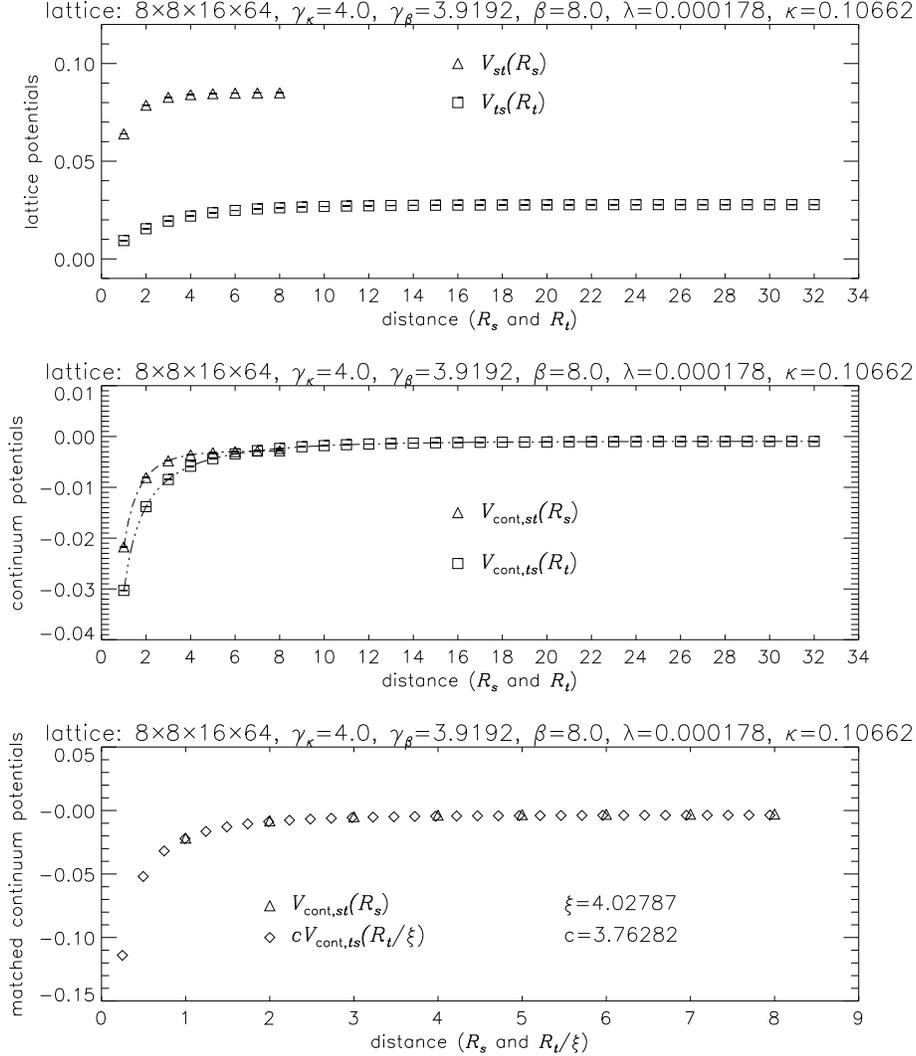,width=12cm}
\caption{\label{PotMPlot} \sl Matching of the (subtracted) lattice 
                              potentials at the perturbative parameters. 
                              The error bars are smaller than the symbols.} 
\end{center}
\end{figure}

The errors of $\xh$, $\xw$ and $\xp$ are relatively large.  
This is caused by the fact that they are determined as ratios of masses with 
individual statistical errors. The jackknife errors quoted are obtained 
from the jackknife samples for the mass ratios themselves.
Calculating the errors from the mass errors using error propagation would 
result in even larger error estimates.
Inspired by a method found in ref.~\cite{BKNS88} one can obtain even smaller 
errors instead, if $\xi$ is directly determined by a matching of
the space- and time-like secondary quantities,
without any reference to the correlation lengths extracted from them
afterwards.
We have realized this proposal for the static potentials in space ($\Vst$)
and time ($\Vts$) direction.
To begin with, we calculated the corresponding continuum potentials
\be
\Vcij(R_i)\equiv\Vij(R_i)-\cij-\dij\gij(\mij,R_i)\,,
\label{contpot}
\ee
since the lattice sum $\iij$ in (\ref{lattsum}) is only meaningful for integer 
$R_i$.
Constant and lattice correction terms in lowest order are found from
eqs.~(\ref{potlatt}) and (\ref{yukawafit}) to be 
\be
\cij+\dij\gij(\mij,R_i)=\frac{3g_R^2}{16\pi}\,\left[\,
\iij(\mij,0)-\iij(\mij,R_i)+\frac{1}{R_i}\,\,\Exp^{\,-\mij R_i}\,\right]\,,
\label{lattpert}
\ee 
while solely in the subtraction step $g_R^2$ and $\mij$ were taken from
table~\ref{PotTab}. 
Hence the matching condition reads
\be
\Vcst(R_s)=c\cdot\Vcts(R_t/\xi)\,,\quad\xp\equiv\xi\,.
\label{ximatch}
\ee
It was fulfilled by fitting the space-like continuum potential to a Yukawa 
shape $-A\,\Exp^{\,-mx}/x+C$ in imitation of (\ref{potclim}), equating the
fit function at arguments $R_t/\xi$ with the time-like potential data times
a constant, and solving every possible equation pair for $\xi$ and $c$.
The final $\xp$--values given in the last row of table~\ref{XiPotTab} are 
averages over all such solutions along that $R_i$--interval, in which the 
two potentials have their characteristic slopes, and interchanging the
r\^oles of $\Vcst$ and $\Vcts$ in eq.~(\ref{ximatch}) always enabled a useful
cross-check. 
As exemplarily reflected in the perturbative simulation parameters on the 
larger lattice in figure~\ref{PotMPlot}, the deviation between the curves 
then becomes uniformly minimal in their whole range.   

The lattice spacing anisotropies from this potential matching
resemble the screening
mass ratios, but the errors from a repetition of this procedure with 1000
normally distributed random data are indeed smaller.
Moreover, $\xp$ is fully compatible with $\xh$ and $\xw$ in the previous
subsection at the perturbative values of the coupling anisotropies.

\newpage
%
%
%
\subsection{Evaluation of the non-perturbative asymmetries and comparison with
the perturbative result}
We have determined the lattice spacing anisotropies $\xi_i$ from Higgs 
($i=H$) and gauge  ($i=W$) boson correlation functions and static potentials
($i=V$) at different pairs of coupling anisotropy parameters.
Since $\gk$ has been held fixed, each $\xi_i$ is looked upon as a function 
of $\gb$, and the requirement of space-time symmetry restoration suggests 
the existence of a unique coupling anisotropy $\uts{\gb}{(np)}$, where all 
$\xi_i$ possess the same value $\uts{\xi}{(np)}$.
This defines the non-perturbative anisotropy parameters. 

\begin{figure}[htb]
\begin{center}
\epsfig{file=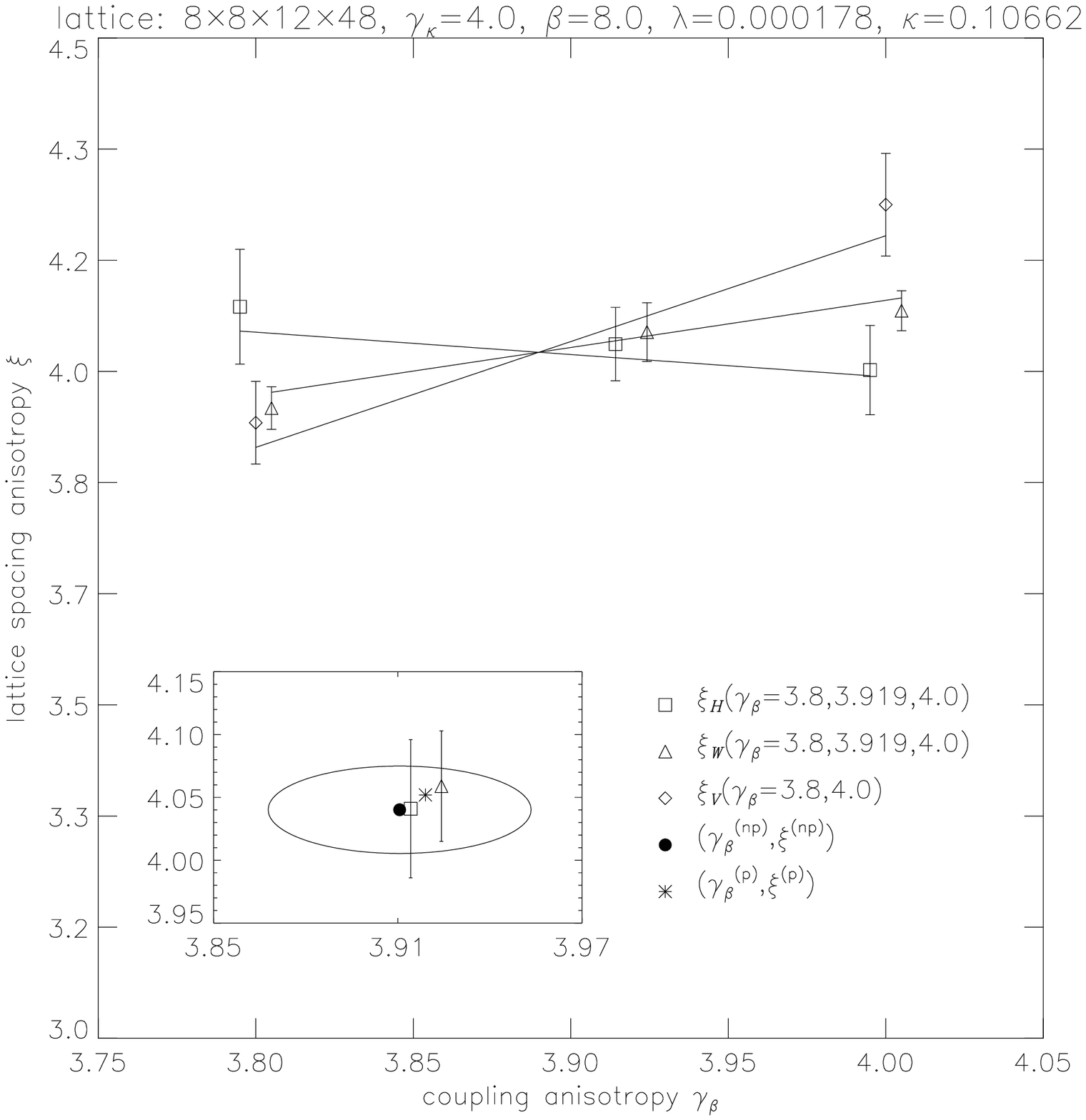,width=8cm}
\epsfig{file=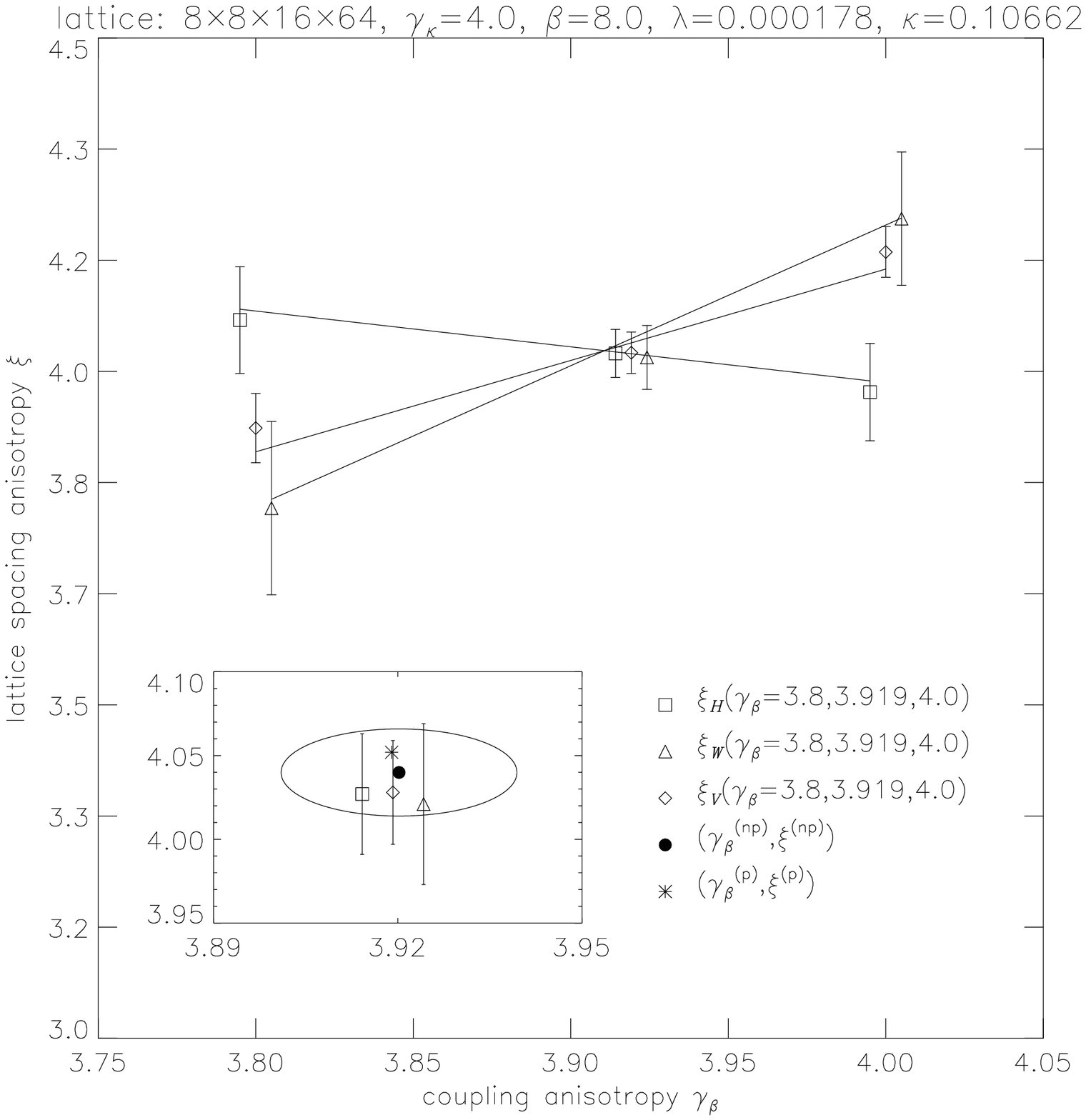,width=8cm}
\caption{\label{XiFinPlot} \sl Final $\xi$--evaluation for both lattices 
                               from the three simulation points,
                               whose equal abscissas are slightly 
                               displaced for better visualization.
                               The inserts show the average matching
                               points and its error ellipses, which enclose 
                               both the numerical estimates at $\gb=3.919$ and 
                               the perturbative result.}
\end{center}
\end{figure}

%
%
Therefore, we linearly interpolate the numbers 
$\xi_{ij}\equiv\xi_i(\gamma_{\beta,j})$ at the three values 
$\gamma_{\beta,j}$ of eq.~(\ref{cplchoice}) within their errors 
$\Delta\xi_{ij}$ to a matching point 
$\big(\uts{\gb}{(np)},\uts{\xi}{(np)}\big)$ by minimizing the sum of squares    
\be
\chi^2=
\sum_i\sum_j\left\{\frac{\xi_{ij}-(\,
\uts{\xi}{(np)}+c_i\,[\,\gamma_{\beta,j}-\uts{\gb}{(np)}\,]\,)}
{\Delta\xi_{ij}}\right\}^2
\label{xiinterp}
\ee
with respect to $c_i$ and the common fit parameters $\uts{\gb}{(np)}$ and 
$\uts{\xi}{(np)}$.
We obtain the final results 
\bea
\mbox{$8^2\times12\times48$ :}& \quad &
\uts{\gb}{(np)}=3.911(43)\,,\quad\uts{\xi}{(np)}=4.040(35) \\
\mbox{$8^2\times16\times64$ :}& \quad &
\uts{\gb}{(np)}=3.920(19)\,,\quad\uts{\xi}{(np)}=4.040(26)
\label{xifinal}
\eea 
with errors coming from 5000 normally distributed random data.
Figure~\ref{XiFinPlot} illustrates that both points agree with the 
simulated $\xi_i$ at the perturbative $\gb$--value as well as with the 
perturbative point itself, and finite-size effects appear to be remarkably 
small.
%
%
It remains to be mentioned that (\ref{xifinal}) includes the $\xp$--values
--- which incidentally were not available at $\gb=3.919$ for the smaller
lattice --- from the matching of the potentials.
Using the weighted averages of the two screening mass ratios in 
table~\ref{XiPotTab} in place of the former, we get the similar results
$\uts{\gb}{(np)}=3.921(38)$, $\uts{\xi}{(np)}=4.038(29)$, 
and $\uts{\gb}{(np)}=3.921(19)$, $\uts{\xi}{(np)}=4.038(26)$, respectively.

All estimates signal a perfect confirmation of the perturbative results
$\uts{\gb}{(p)}=3.919$ and $\uts{\xi}{(p)}=4.052$ calculated in section II. and 
quoted in item e.  at the end of subsection II.B.
There is no evidence that the unknown higher-order corrections in $g^2$
and $\lambda$ could lead to any visible modifications, which would make the 
applicability of one-loop perturbation theory to the anisotropy parameters 
doubtful.
In conclusion, the non-perturbative contributions can not be resolved within
the intrinsic errors of numerical simulations, and as a consequence, the 
perturbative choice of the anisotropy parameters in investigations with the
 SU(2)-Higgs model with asymmetric lattice parameters is justified
 also for Higgs masses $\mh\ge 80$ GeV.

\section{Summary }

In summary, we have worked out the complete one-loop perturbation theory of 
the SU(2)-Higgs model on lattices with asymmetric lattice spacings in 
$R_\xi$ gauges. 
We have determined the critical hopping parameter and the coupling 
asymmetries in one-loop perturbation theory, as a function of the asymmetry 
parameter $\xi$. We have proven by explicit calculations the gauge 
independence of these results in $R_\xi$ gauges. 
We have perturbatively studied the approach to the 
continuum limit of the finite temperature theory and have determined the 
optimal choice of $\xi$ ensuring the most economical lattice simulation for a 
given precision determination of the physical parameters. 

To test the relevance of the perturbative results to non-perturbative 
studies we have determined the non-perturbative coupling anisotropies 
using lattice simulations. Three channels have been studied, namely 
Higgs and W--masses as well as the static potential. For our parameters, 
i.e. Higgs mass near 80 GeV,    $g^2 \approx 0.5$ and 
$\xi=4$ the perturbative results agree with the 
non-perturbative determination within the (high) accuracy of  the latter. 
This result opens the possibility to perform 
lattice simulation using the perturbative coupling anisotropies 
without the need of a non-perturbative determination.
In particular our results are essential to study the electroweak
 phase transition for Higgs boson masses around or above 80 GeV and
determine the properties of the hot electroweak plasma.

\vspace{.5cm}

This work was partially supported by
Hungarian Science Foundation grant under Contract  No. OTKA-T016248 and 
OTKA-T022929 and by Hungarian Ministry of Education 
grant No. FKFP-0128/1997.

\vfill


\newpage

\begin{thebibliography}{99}
\bibitem{kuzmin}
V. A. Kuzmin, V. A. Rubakov, M. E. Shaposhnikov, Phys. Lett. {\bf B155},
 36 (1985).
\bibitem{arnold}
P. Arnold, O. Espinosa, Phys. Rev. {\bf D47}  3546 (1993);
W. Buchm\"uller, Z. Fodor, T. Helbig ,   D. Walliser,
Ann. Phys. {\bf 234}  260 (1994).
\bibitem{fodor}
Z. Fodor ,   A. Hebecker, Nucl. Phys. {\bf B432}  127 (1994).
\bibitem{buchmuller}
W. Buchm\"uller, Z. Fodor, A. Hebecker, Nucl. Phys. {\bf B447} 317 (1995).
\bibitem{bunk}
B. Bunk, E.-M. Ilgenfritz, J. Kripfganz ,   A. Schiller,
Nucl. Phys. {\bf B403}  453 (1993).
\bibitem{kajantie}
K. Farakos, K. Kajantie, K. Rummukainen, M. Shaposhnikov,
Nucl. Phys. {\bf B407} 356 (1993); {\bf B425} 67 (1994); {\bf B442} 317 (1995);
K. Kajantie, M. Laine, K. Rummukainen, M. Shaposhnikov,
Nucl.Phys. {\bf B466}  189 (1996).
\bibitem{laine} M. Laine, Nucl. Phys. {\bf B451}  484 (1995).
\bibitem{karsch}
F. Karsch, T. Neuhaus ,   A. Patk\'os, Nucl. Phys. {\bf B441}  629 (1995).
\bibitem{patkos} A. Jakov\'ac, K. Kajantie, A. Patk\'os,
Phys. Rev. {\bf D49}  6810 (1994);
A. Jakov\'ac, A. Patk\'os, P. Petreczky, Phys.Lett. {\bf B367}  283 (1996).
\bibitem{heidelberg} H.-G. Dosch, J. Kripfganz, A. Laser,
M. G. Schmidt, Phys. Lett. {\bf B365}  213 (1995); E. M. Ilgenfritz,
J. Kripfganz, H. Perlt, A. Schiller, Phys. Lett. {\bf B356}
 561 (1995).
\bibitem{philipsen}
W. Buchm\"uller, O. Philipsen, Nucl. Phys.
{\bf B443}  47 (1995); O. Philipsen,
M. Teper, H. Wittig, Nucl.Phys. {\bf B469}  445 (1996).
\bibitem{ape}
F. Csikor, Z. Fodor, J. Hein, K. Jansen, A. Jaster, I. Montvay,
Phys. Lett. {\bf B334}  405 (1994);
F. Csikor, Z. Fodor, J. Hein ,   J. Heitger,
Phys. Lett. {\bf B357}  156 (1995).
\bibitem{FHJJM95}
Z. Fodor, J. Hein, K. Jansen, A. Jaster, I. Montvay,
Nucl. Phys. {\bf B439}  147 (1995).
%
\bibitem{CFHJM96}
F. Csikor, Z. Fodor, J. Hein, A. Jaster, I. Montvay,
Nucl. Phys. {\bf B474}  421 (1996).
\bibitem{B95}
Z. Fodor, K. Jansen, Phys. Lett. {\bf B331}  119 (1994); B. Bunk,
Nucl. Phys. {\bf B42} (Proc. Suppl.)  566 (1995).
%
\bibitem{jansen} K. Jansen, Nucl. Phys. {\bf B47} (Proc. Suppl.)  196 (1996); 
K. Rummukainen, Nucl. Phys. {\bf B53}  (Proc. Suppl.) 30 (1997).
\bibitem{CFHHJM97}
F. Csikor, Z. Fodor, J. Hein, J. Heitger, A. Jaster, I. Montvay,
Nucl. Phys. {\bf B53} (Proc. Suppl.)  612 (1997).
%
\bibitem{shap} K. Kajantie, M. Laine, K. Rummukainen, M. Shaposhnikov,
Phys. Rev. Lett. {\bf 77}  2887 (1996).
\bibitem{Leipzig} F. Karsch, T. Neuhaus, A. Patk\'os, J. Rank,
Nucl. Phys. {\bf B53} (Proc. Suppl.)  623 (1997);
M. G\"urtler, E.-M. Ilgenfritz,
A. Schiller, Phys. Rev. {\bf D56}  3888 (1997).
\bibitem{bender} I. Bender et al., Nucl. Phys. {\bf B17} (Proc. Suppl.)
  387 (1990); B. Bunk, ibid {\bf B42}  566 (1995).
\bibitem{karsch82}
F. Karsch, Nucl. Phys. {\bf B205}  285 (1982).
\bibitem{karsch89}
F. Karsch, O. Stamatescu, Phys. Lett. {\bf B227}  153 (1989).
\bibitem{Cs-F}
F. Csikor, Z. Fodor, Phys. Lett. {\bf B380}  113 (1996).
\bibitem{montvay}
I. Montvay, Phys. Lett. {\bf B172} (1986) 71; Nucl. Phys. {\bf B293}  479 (1987)
.
\bibitem{MMu} I. Montvay, G. M\"unster, Quantum fields on a lattice
Cambridge Univ. Press, Cambridge (1994)
\bibitem{hasenfratz} A. Hasenfratz, P. Hasenfratz, Phys. Rev.
{\bf D34}  3160 (1986).
%
\bibitem{Wolff} U. Wolff, Phys. Lett. {\bf B288}  166 (1992);
 K. Akemi et al., Phys. Lett. {\bf B328}  407 (1994).
\bibitem{HH96}
J. Hein, J. Heitger,
Phys. Lett. {\bf B385}  242 (1996).
%
\bibitem{CPT84}
G. Curci, G. Pafutti, R. Tripiccione,
Nucl. Phys. {\bf B240} [FS12]  91 (1984).
\bibitem{HK85}
U. Heller, F. Karsch,
Nucl. Phys. {\bf B251} [FS13]  254 (1985).
%
\bibitem{LMP86}
W. Langguth, I. Montvay, P. Weisz,
Nucl. Phys. {\bf B277}  11 (1986).
%
\bibitem{S94}
R. Sommer,
Nucl. Phys. {\bf B411}  839 (1994).
%
\bibitem{M94MMc95}
C. Michael,
Phys. Rev. {\bf D49}  2616 (1994);
C. Michael, A. McKerrel,
Phys. Rev. {\bf D51}  3745 (1995).
%
\bibitem{BKNS88}
G. Burgers, F. Karsch, A. Nakamura, I.O. Stamatescu,
Nucl. Phys. {\bf B304}  587 (1988).
%

%
%
\end{thebibliography}
\end{document}